\documentclass[
aps,nofootinbib,showpacs,showkeys,preprint,tightenlines ] {revtex4}  

\usepackage{graphicx}
\usepackage{epsfig,subfig}
\usepackage{amsmath,amssymb}
\usepackage{color}

\newcommand{\dis}[1]{\begin{equation}\begin{split}#1\end{split}\end{equation}}
\newcommand{\be}{\begin{equation}}
\newcommand{\ee}{\end{equation}}
\def\bea{\begin{eqnarray}}
\def\eea{\end{eqnarray}}

\newcommand{\eq}[1]{Eq.~(\ref{#1})}
\newcommand{\eqs}[1]{Eqs.~(\ref{#1})}

\newcommand{\cblue}[1]{{\color{blue}{#1}}}

\newcommand{\VEV}[1]{\langle #1 \rangle}

\newcommand{\tev}{\,\textrm{TeV}}
\newcommand{\gev}{\,\textrm{GeV}}

\def\tb{\tan\beta}

\begin{document}

\title{\Large\bf 
Naturalness-guided Gluino Mass Bound 
\\from the Minimal Mixed Mediation of SUSY Breaking
}

\author{Doyoun Kim$^{(a)}$\footnote{email: doyoun.kim@apctp.org}
and Bumseok Kyae$^{(b)}$\footnote{email: bkyae@pusan.ac.kr}
}
\affiliation{
$^{(a)}$ Asia Pacific Center for Theoretical Physics, Pohang, Gyeongbuk 790-784, Korea
\\
$^{(b)}$ Department of Physics, Pusan National University, Busan 609-735, Korea
}

\begin{abstract}

In order to significantly reduce the fine-tuning  
associated with the electroweak symmetry breaking 
in the minimal supersymmetric standard model (MSSM), 
we consider  
not only the minimal gravity mediation effects 
but also the minimal gauge mediation ones 
for a common supersymmetry breaking source 
at a hidden sector.   
In this ``Minimal Mixed Mediation model,'' 
the minimal forms for the K${\rm \ddot{a}}$hler potential 
and the gauge kinetic function 
are employed at tree level. 
The MSSM gaugino masses are radiatively generated 
through the gauge mediation. 
Since a ``focus point'' of the soft Higgs mass parameter, 
$m_{h_u}^2$ appears around $3$-$4 \tev$ energy scale 
in this case,  
$m_{h_u}^2$ is quite insensitive to stop masses. 
Instead, the naturalness of the small $m_{h_u}^2$ 
is more closely associated with the gluino mass  
rather than the stop mass
unlike the conventional scenario. 
As a result, even a $3$-$4 \tev$ stop mass, 
which is known to explain the $125 \gev$ Higgs mass 
at three-loop level, 
can still be compatible with the naturalness of 
the electroweak scale.  
On the other hand, 
the requirements of various fine-tuning measures 
much smaller than 100 and $|\mu|<600 \gev$ 
constrain the gluino mass to be      
$1.6 \tev \lesssim m_{\tilde{g}} \lesssim 2.2 \tev$, 
which is well inside the discovery potential range 
of LHC Run~II.

\end{abstract}



\maketitle


\section{Introduction}

How to naturally keep the small Higgs boson mass 
against its quadratically divergent radiative corrections
has been one of the most important issues in the particle physic community 
for the last four decades. 
Since this question raised in the Standard Model (SM) 
is associated with 
stabilization of the EW scale against the grand unified 
theory (GUT) scale or the Planck scale, 
many ideas and theories beyond the SM and 
towards the fundamental theory 
have been motivated and suggested in order to address this question. 
The supersymmetric (SUSY) resolution to it is 
to cancel the quadratic divergences 
by introducing superpartners with spins different by $1/2$
from those of the SM particles, and 
their interactions with the same strength as those of the SM.
All of them can consistently be controlled 
within the SUSY framework \cite{book}.  

Since the top quark and its superpartner ``stop'' 
dominantly contribute to the radiative Higgs mass 
via the large top quark Yukawa coupling, 
the stop mass has been regarded as a barometer 
for naturalness of the minimal SUSY SM (MSSM): 
a stop mass lighter than $1 \tev$ is quite essential
for keeping the naturalness of the EW scale and the Higgs boson mass. 
However, the experimental mass bound on the stop has already 
exceeded $700 \gev$ \cite{stopmass}. 
Thus, it would be very timely to ask  
whether the low energy SUSY can still remain natural 
even with a somewhat heavy stop mass 
greater than $1 \tev$.   

On the other hand, the gluino is not directly involved 
in this issue, because it does not couple 
to the Higgs boson at tree level. 
Instead, the gluino mass dominantly influences 
the renormalization group (RG) evolution of the stop mass parameters. 
In this sense, the gluino affects the Higgs mass parameter $m_{h_u}^2$ just indirectly in the ordinary MSSM. 
In this paper, however, we attempt to investigate another possibility: 
the gluino can play a more important role 
in the naturalness of the small Higgs boson mass. 
As a consequence, the stop mass can be much less responsible 
for it: it can be much heavier 
than the present experimental bound.   
Indeed, the gluino can be more easily explored than the stop 
at the Large Hadron Collider (LHC).         
Thus, if a relatively light gluino mass turns out 
to be needed,  
this scenario could readily be tested at LHC Run~II. 



Because of the top quark Yukawa coupling constant 
$y_t$ of order unity, as mentioned above,  
the top quark and stop make the dominant contributions 
not only to the renormalization of a soft mass parameter 
of the Higgs $h_u$ ($\equiv\Delta m_{h_u}^2$),  
but also to the radiative physical Higgs mass 
($\equiv\Delta m_H^2$) \cite{book,twoloop}:
\bea \label{renormHiggs}
&&  ~~\Delta m_{h_u}^2|_{\rm 1-loop}\approx \frac{3|y_t|^2}{8\pi^2}\widetilde{m}_t^2 
~{\rm log}\left(
\frac{\widetilde{m}_t^2}{\Lambda^2}\right)
\left[1 + \frac12\frac{A_t^2}{\widetilde{m}_t^2}\right] ,
\\ \label{physicalHiggs}
&&\Delta m_H^2|_{\rm 1-loop}\approx 
\frac{3m_t^4}{4\pi^2v_h^2}\left[{\rm log}\left(\frac{\widetilde{m}_t^2}{m_t^2}\right) 
+\frac{A_t^2}{\widetilde{m}_t^2}\left(1-\frac{1}{12}\frac{A_t^2}{\widetilde{m}_t^2}\right)\right] ,
\eea
where $m_t$ ($\widetilde{m}_t$) denotes the top quark (stop) mass, and $v_h$ 
is the vacuum expectation value (VEV) of the Higgs boson, $v_h\equiv\sqrt{\VEV{h_u}^2+\VEV{h_d}^2}\approx 174~{\rm GeV}$  with $\tan\beta\equiv\VEV{h_u}/\VEV{h_d}$. 
For simplicity, here we assumed that 
the SU(2)$_L$-doublet and -singlet stops 
(``LH- and RH stops'') are degenerate, and 
the ``$A$-term'' coefficient corresponding to the top quark Yukawa coupling,  
$A_t$ dominates over $\mu \cdot{\rm cot}\beta$, 
where $\mu$ is the ``Higgsino'' mass. 
By introducing SUSY, thus, the quadratic dependence on 
the ultraviolet (UV) cutoff $\Lambda$ in the SM 
for $\Delta m_{h_u}^2|_{\rm 1-loop}$
is replaced by a logarithmic one as seen in \eq{renormHiggs}.  
For a small enough $\Delta m_{h_u}^2|_{\rm 1-loop}$, however, 
the stop mass should necessarily be small enough. 
Otherwise, the Higgs mass parameters, $m_{h_u}^2$ and $m_{h_d}^2$, should be finely tuned with $\mu$ to yield  
the measured value of the $Z$ boson mass $m_Z\approx 91 \gev$,  
because they are related to each other 
via the minimization condition of the Higgs potential \cite{book}, 
\dis{ \label{m_Z}
\frac12 m_Z^2=\frac{m_{h_d}^2-m_{h_u}^2{\rm tan}^2\beta}{{\rm tan}^2\beta-1}
-|\mu|^2 . 
}

As seen in \eq{physicalHiggs}, the radiative correction to 
the physical Higgs mass depends logarithmically on the stop mass. 
Actually the tree level Higgs mass in the MSSM 
should be lighter even than the $Z$ boson mass 
($<m_Z\cdot{\rm cos}2\beta$) \cite{book}. 
Thus, the radiative Higgs mass \eq{physicalHiggs} is also quite  
essential for explaining the observed Higgs boson mass. 
In view of \eq{physicalHiggs}, however, the recently measured 
Higgs boson mass, $125 \gev$ \cite{LHCHiggs} 
is indeed too heavy as a SUSY Higgs mass, 
because it would require a too heavy stop mass 
(``little hierarchy problem''). 
Many SUSY models have been proposed 
for raising the Higgs boson mass by extending the MSSM, 
but still assuming a relatively light stop, 
$\widetilde{m}_t\lesssim 1 \tev$ \cite{Kyae}.
However, the experimental mass bound on the stop has already 
exceeded $700 \gev$ \cite{stopmass}, as mentioned above. 
Of course, the second term in \eq{physicalHiggs} 
could be helpful for raising the Higgs mass, 
when it is almost maximized, 
$A_t^2/\widetilde{m}_t^2\approx 6$ \cite{book,twoloop}. 
But it is not easy to realize at low energies 
from a UV model via its RG running, 
unless we suppose a tachyonic stop at the GUT scale ($M_G$) \cite{HDK}.   

According to the recent analysis based 
on three-loop calculations in Ref.~\cite{3-loop}, 
a $3$-$4 \tev$ stop mass can account for 
the $125 \gev$ Higgs boson mass with ignorable $A_t$ terms.  
Such a heavy stop mass would give rise to 
a more serious fine-tuning problem associated with 
the light $Z$ boson mass as seen in \eqs{renormHiggs} 
and (\ref{m_Z}), particularly, 
when the cutoff scale $\Lambda$ is 
about GUT scale ($\sim 10^{16} \gev$): 
apparently a fine-tuning of order $10^{-4}$ (or $\Delta_{m_0^2}\sim 10^{+4}$ 
in terms of the fine-tuning measure defined later)
looks unavoidable in the MSSM. 
%
%
To more precisely discuss the UV dependence of $m_{h_u}^2$, addressing the little hierarchy problem, however, 
one should analyze the full RG equations 
under a given specific UV model. 
If a SUSY UV model turns out to be simple enough, 
addressing the above question, 
SUSY could still be recognized as an attractive solution 
to the gauge hierarchy problem.    

A potentially promising UV model is the ``focus point (FP) scenario'' \cite{FMM1}. 
Since it is based on the minimal gravity mediation (mGrM) of 
SUSY breaking, all the soft squared masses including 
the two Higgs mass parameters $m_{h_u}^2$ and $m_{h_d}^2$, 
LH- and RH stop's squared masses $m_{q_3}^2$ and $m_{u_3^c}^2$, etc. 
as well as the MSSM gaugino masses take the universal forms \cite{book,PR1984}:
\dis{
m_{h_u}^2=m_{h_d}^2=m_{q_3}^2=m_{u_3^c}^2=\cdots\equiv m_0^2  
\quad {\rm and} \quad M_3=M_2=M_1\equiv m_{1/2} ,  
}  
where 
$M_{3,2,1}$ denote the gluino, wino, and bino masses, respectively. 
In this case, as noticed in Ref.~\cite{FMM1}, 
the RG flows of $m_{h_u}^2$ converge 
about the $Z$ boson mass scale to a small negative value, 
regardless of its initial values taken at the GUT scale, 
i.e., various $m_0^2$ values,   
only if the $A_t$ and $m_{1/2}$ are sufficiently suppressed.  
Since $m_{h_u}^2$ is almost independent of $m_0^2$, 
a small enough $m_{1/2}$ turns out to be 
responsible for a small negative $m_{h_u}^2$, 
naturally explaining the smallness of the EW scale or $m_Z$ compared to the GUT or Planck scale.  
Such a parameter choice can indeed reduce the fine-tuning considerably.   
Several different definitions of the fine-tuning report a similar tendency 
around the ``FP region'' in the MSSM parameter space~\cite{DYK}.    
On the other hand, the low energy values of 
other soft mass parameters 
such as $m_{q_3}^2$ and $m_{u_3^c}^2$ are very sensitive to 
$m_0^2$ values.  
These features in the mGrM might open a possibility to  
naturally explain the smallness of $m_{h_u}^2$ 
in contrast to large stop mass parameters.  

However, the experimental gluino mass bound has already 
exceeded $1.3 \tev$ \cite{gluinomass}, 
and so the unified gaugino mass $m_{1/2}$ 
cannot be small any longer. 
Also the naturalness on a small $A$-term would be questionable. 
Most of all, if the stop masses are around $3$-$4 \tev$, 
they should decouple below the $3$-$4 \tev$ energy scale 
from the ordinary MSSM RG equations, 
and so the FP behavior of $m_{h_u}^2$ becomes 
seriously spoiled below the stop mass scale \cite{KS}. 
Basically the FP scale in the mGrM 
is too far below the stop mass scale desired for 
explaining the $125 \gev$ Higgs boson mass. 
All such problems in the FP scenario arise 
because heavier masses for the Higgs, stop, and gluino 
are experimentally and/or theoretically compelled.   

The best resolution to such problems would be 
to somehow push the FP scale from the $Z$ boson mass scale 
to the desired stop mass scale (``shifted FP'' \cite{BK}) 
such that the $m_0^2$ dependence of $m_{h_u}^2$ becomes 
suppressed before stops are decoupled 
from the RG equation of $m_{h_u}^2$ \cite{KS,BK}. 
Actually, it is indispensable for restoring the naturalness 
of the low energy SUSY in the framework of the FP scenario.
$m_{h_u}^2$ below the stop mass scale or at the $Z$ boson mass scale
can be estimated 
using the Coleman-Weinberg potential \cite{book,CQW}:
\bea \label{RGsm}
m_{h_u}^2(m_Z)&\approx& m_{h_u}^2(\Lambda_T) + 
\frac{3|y_t|^2}{16\pi^2}\bigg[
\sum_{i=q_3,u_3^c}m_{i}^2\left\{{\rm log}\frac{m_{i}^2}{\Lambda_T^2}-1\right\}
-2m_t^2\left\{{\rm log}\frac{m_t^2}{\Lambda_T^2}-1\right\}\bigg]\Bigg|_{\Lambda_T}
\nonumber \\
&\approx& m_{h_u}^2(\Lambda_T) - \frac{3|y_t|^2}{16\pi^2}
\bigg\{m_{q_3}^2+m_{u_3^c}^2\bigg\}
\bigg[1-\frac{m_{q_3}^2-m_{u_3^c}^2}
{2(m_{q_3}^2+m_{u_3^c}^2)}
~{\rm log}\frac{m_{q_3}^2}
{m_{u_3^c}^2}\bigg]\Bigg|_{\Lambda_T} , 
\eea  
where the cutoff $\Lambda_T$ is set to the stop decoupling scale 
($\approx \sqrt{m_{q_3}m_{u_3^c}}$). 
The last term of the second line in \eq{RGsm} is relatively 
suppressed.  
Since the $m_0^2$ dependence of stop masses would be loop suppressed,   
$m_{h_u}^2$ needs to be well focused around $\Lambda_T$. 
Due to the additional negative contribution to $m_{h_u}^2(m_Z)$ below $\Lambda_T$, 
a small positive $m_{h_u}^2(\Lambda_T)$ would be more desirable.  

In order to push the FP scale up to 
the desired stop mass scale, $3$-$4 \tev$,  
we will consider the gauge mediation effects as well as 
the mGrM effects
for a common SUSY breaking source at the hidden sector, 
introducing some messenger fields: 
we will attempt to combine the two representative SUSY breaking 
mediation scenarios, the mGrM and 
the minimal gauge mediation (mGgM) at the GUT scale 
in a single supergravity (SUGRA) framework \cite{BK}.  
We call it the ``Minimal Mixed Mediation'' of SUSY breaking.
%
%
For a qualitative understanding on the FP behaviors, 
in this paper we will present the semianalytic solutions 
to the relevant RG equations for small $\tb$ cases. 
Also we will perform their full numerical analyses  
for large $\tb$ cases. 
Based on these results, we will explore the parameter space  
that can naturally explain the small Higgs mass parameter, 
and then derive the gluino mass bound consistent with it.

This paper is organized as follows: 
in Section~\ref{sec:RGsol} we will present 
semianalytic RG solutions for $m_{h_u}^2$ and the stop masses 
in the MSSM with a small $\tb$. 
They will be utilized in the subsequent sections.   
We will leave the details of their derivations  
in the Appendix.   
In Section~\ref{sec:mGrM}, we will discuss 
why the fine-tunings become more serious in the mGrM 
with relatively heavy stop masses.  
In Section~\ref{sec:MMM}, we will introduce 
the Minimal Mixed Mediation of SUSY breaking 
and show that it significantly reduces the fine-tunings 
of the MSSM. 
In this section, we will derive a proper gluino mass bound 
consistent with the naturalness of the EW scale 
and the Higgs boson mass. 
Section~\ref{sec:conclusion} will be devoted to the Conclusion. 


\section{Semianalytic RG solutions} \label{sec:RGsol}

In this section, we will first present 
our semianalytic solutions 
to the RG equations of some soft SUSY breaking mass parameters 
in small $\tb$ cases. 
When $\tb$ is large, the expressions on them are not simple 
enough, and so one should perform a full numerical analysis. 
As will be seen later, however, large $\tb$ cases 
turn out to be much more useful for 
reducing the fine-tuning of the EW scale. 
Nonetheless, discussions on the small $\tb$ case 
would be helpful for a qualitative understanding on 
the structure of the FP of $m_{h_u}^2$ and 
for getting an intuition on how to resolve the problem.

When ${\rm tan}\beta$ is small enough and the RH neutrinos are decoupled (by assuming their small Yukawa couplings), 
the RG evolutions of the soft mass parameters, 
$m_{q_3}^2$, $m_{u^c_3}^2$, $m_{h_u}^2$, and $A_t$ are described with the following simple equations \cite{book}:  
\begin{eqnarray}
16\pi^2\frac{dm_{q_3}^2}{dt}&=&2y_t^2\left(X_t+A_t^2\right)-\frac{32}{3}g_3^2M_3^2-6g_2^2M_2^2-\frac{2}{15} g_1^2M_1^2 ,
\label{RG1} \\
16\pi^2\frac{dm_{u^c_3}^2}{dt}&=&4y_t^2\left(X_t+A_t^2\right)-\frac{32}{3}g_3^2M_3^2-\frac{32}{15} g_1^2M_1^2 ,
\label{RG2} \\
16\pi^2\frac{dm_{h_u}^2}{dt}&=&6y_t^2\left(X_t+A_t^2\right)-6g_2^2M_2^2-\frac65 g_1^2M_1^2 ,
\qquad~~ \label{RG3} \\
8\pi^2\frac{dA_t}{dt}&=&6y_t^2A_t-\frac{16}{3}g_3^2M_3-3g_2^2M_2-\frac{13}{15} g_1^2M_1 ,
\label{RG4}
\end{eqnarray} 
where $t$ parametrizes the renormalization scale $Q$, 
$t-t_0={\rm log}\frac{Q}{M_{G}}$, and $X_t$ is defined as 
$m_{q_3}^2+m_{u^c_3}^2+m_{h_u}^2$. 
Here we neglected the bottom quark Yukawa coupling $y_b$, 
the sbottom quark's squared mass $m_{d_3^c}^2$, 
and also the leptonic contributions 
due to the smallness of $\tb$. 
In the above equations, 
the RG evolutions for the MSSM gauge couplings $g_{3,2,1}$ and 
the gaugino masses $M_{3,2,1}$ are already well known \cite{book}: 
\begin{eqnarray} \label{gaugeSol}
g_a^2(t)=\frac{g_0^2}{1-\frac{g_0^2}{8\pi^2}b_a(t-t_0)}~,
~~~~ {\rm and}~~~~~ 
\frac{M_a(t)}{g_a^2(t)}=\frac{m_{1/2}}{g_0^2}~ ,
\end{eqnarray}
where $g_0$ and $m_{1/2}$ denote the unified gauge coupling 
constant and the unified gaugino mass, respectively, and   
$b_a$ ($a=3,2,1$) means the beta function coefficients 
for the MSSM field contents, 
$(b_3,b_2,b_1)=(-3,1,\frac{33}{5})$.
For the full RG equations valid when $\tb$ is large, 
refer to the Appendix of Ref.~\cite{KS}. 
The semianalytic solutions for $m_{q_3}^2$, $m_{u^c_3}^2$, 
and $m_{h_u}^2$ turn out to take the following forms:
\begin{eqnarray}
&&m_{q_3}^2(t)=m_{q_30}^2+\frac{X_0}{6}\left[e^{\frac{3}{4\pi^2}\int^{t}_{t_0}dt^\prime y_t^2}-1\right]
+\frac{F(t)}{6}
\label{Sol1} \\ \nonumber 
&&\qquad +\left(\frac{m_{1/2}}{g_0^2}\right)^2
\left[\frac89\bigg\{g_3^4(t)-g_0^4\bigg\}
-\frac32\bigg\{g_2^4(t)-g_0^4\bigg\}-\frac{1}{198}\bigg\{g_1^4(t)-g_0^4\bigg\}\right] ,
\\
%
&&m_{u^c_3}^2(t)=m_{u^c_30}^2+\frac{X_0}{3}\left[e^{\frac{3}{4\pi^2}\int^{t}_{t_0}dt^\prime y_t^2}-1\right]
+\frac{F(t)}{3}
\label{Sol2} \\ \nonumber 
&&\qquad +\left(\frac{m_{1/2}}{g_0^2}\right)^2
\left[\frac89\bigg\{g_3^4(t)-g_0^4\bigg\}
-\frac{8}{99}\bigg\{g_1^4(t)-g_0^4\bigg\}\right] , 
\\
&&m_{h_u}^2(t)=m_{h_u0}^2+\frac{X_0}{2}\left[e^{\frac{3}{4\pi^2}\int^{t}_{t_0}dt^\prime y_t^2}-1\right]
+\frac{F(t)}{2}
\label{Sol3} \\ \nonumber 
&&\qquad -\left(\frac{m_{1/2}}{g_0^2}\right)^2 
\left[\frac32\bigg\{g_2^4(t)-g_0^4\bigg\}
+\frac{1}{22}\bigg\{g_1^4(t)-g_0^4\bigg\}\right] ,
\end{eqnarray}
where the subscript $0$ in $m_{q_30}^2$, $m_{u^c_30}^2$, $m_{h_u0}^2$, and $X_0$
($\equiv m_{q_30}^2+m_{u^c_30}^2+m_{h_u0}^2$)   
means the values of the corresponding mass parameters 
at the GUT scale, or $t=t_0\equiv {\rm log}(M_G/{\rm GeV})$. 
In these solutions, $F(t)$ is given by 
\dis{ \label{F(t)}
&F(t)\equiv 
\frac{1}{64\pi^4}\left(\frac{m_{1/2}}{g_0^2}\right)^2\bigg[
\left(e^{\frac{3}{4\pi^2}\int^{t}_{t_0}dt^\prime y_t^2}\int^{t}_{t_0}dt^\prime ~G_A ~e^{\frac{-3}{4\pi^2}\int^{t^\prime}_{t_0}dt^{\prime\prime} y_t^2}\right)^2
\\
&\qquad \qquad\qquad\qquad\quad -2 ~e^{\frac{3}{4\pi^2}\int^{t}_{t_0}dt^\prime y_t^2}
\int^{t}_{t_0}dt^\prime ~G_A \int^{t^\prime}_{t_0}dt^{\prime\prime}
~G_A~
e^{\frac{-3}{4\pi^2}\int^{t^{\prime\prime}}_{t_0}dt^{\prime\prime\prime} y_t^2}\bigg]
\\
&\quad -\frac{1}{4\pi^2}\left(\frac{m_{1/2}}{g_0^2}\right)^2
\left[e^{\frac{3}{4\pi^2}\int^{t}_{t_0}dt^\prime y_t^2} \int^{t}_{t_0}dt^\prime ~G_X^2 ~e^{\frac{-3}{4\pi^2}\int^{t'}_{t_0}dt^{\prime\prime}y_t^2}
-\int^{t}_{t_0}dt^\prime~G_X^2 \right]
\\
&\quad +\frac{A_0}{4\pi^2}\left(\frac{m_{1/2}}{g_0^2}\right) ~e^{\frac{3}{4\pi^2}\int^{t}_{t_0}dt^\prime y_t^2}\left[
\int^{t}_{t_0}dt^\prime ~G_A-e^{\frac{3}{4\pi^2}\int^{t}_{t_0}dt^\prime y_t^2} \int^{t}_{t_0}dt^\prime ~G_A ~e^{\frac{-3}{4\pi^2}\int^{t^\prime}_{t_0}dt^{\prime\prime} y_t^2}
\right]
\\
&\quad +A_0^2~e^{\frac{3}{4\pi^2}\int^{t}_{t_0}dt^\prime y_t^2}
\bigg[e^{\frac{3}{4\pi^2}\int^{t}_{t_0}dt^\prime y_t^2}
-1\bigg] ,  
%
}
where $A_0\equiv A_t(t=t_0)$, and 
$G_A$ and $G_X^2$ are defined as 
\dis{ \label{G_A}
G_A(t)\equiv 
\left[\frac{16}{3}g_3^4(t)+3g_2^4(t)+\frac{13}{15}g_1^4(t)\right] 
~~ {\rm and} ~~
G_X^2(t)\equiv 
\left[\frac{16}{3}g_3^6(t)+3g_2^6(t)+\frac{13}{15}g_1^6(t)\right] ,
}
respectively. 
For details of the above solutions, refer to the Appendix. 
Numerical calculation shows that the sign of $F(t)$ is negative, and 
$|F(t)/2|$ is larger than the second line of \eq{Sol3}, 
which is positive.  
Consequently larger values of $(m_{1/2}/g_0^2)$ and $A_0$ 
lead to large negative values of $m_{h_u}^2$ at low energies \cite{KS}. 

The initial values, $m_{q_30}^2$, $m_{u^c_30}^2$, and $m_{h_u0}^2$ should be determined by a UV model. 
They would be associated with a SUSY breaking mechanism.  
We will discuss it in the following sections.

\section{Minimal Gravity Mediation} \label{sec:mGrM}

The FP scenario is based on the mGrM model. 
In this section, we will first review the mGrM of 
SUSY breaking, 
particularly investigating the UV boundary conditions 
on the relevant soft mass parameters, 
and then discuss the FP in the mGrM model.

\subsection{Basic Setup in the Minimal Gravity Mediation}

The $N=1$ SUGRA Lagrangian is described basically with 
the K${\rm{\ddot{a}}}$hler potential $K$, 
superpotential $W$, and gauge kinetic function $f_{ab}$. 
In the mGrM scenario or minimal SUGRA (mSUGRA) model, 
particularly, the minimal form of 
the K${\rm{\ddot{a}}}$hler potential is employed, and 
the superpotentials of the hidden and observable sectors 
are separated:   
\dis{ \label{KSpot} 
K=\sum_{i} |z_i|^2+\sum_{r}|\phi_r|^2 ~, \quad
W=W_H(z_i)+W_O(\phi_r)
}
where $z_i$ ($\phi_r$) denotes scalar fields in the hidden (observable) sector. 
The kinetic terms of $z_i$ and $\phi_r$, hence, have the canonical form. 
For the hidden sector scalar fields $z_i$s, and 
the hidden sector superpotential $W_H$, 
nonzero VEVs are assumed\cite{PR1984}: 
\dis{ \label{vev}
\langle z_i\rangle=b_iM_P , ~ 
\langle\partial_{z_i} W_H\rangle=a_i^*mM_P , ~ 
\langle W_H\rangle=mM_P^2 ,
}
where $a_i$ and $b_i$ are dimensionless numbers 
and $M_P$ ($\approx 2.4\times 10^{18} \gev$) 
means the reduced Planck mass. 
Then, $\langle W_H\rangle$ or $m$ yields the gravitino mass, 
$m_{3/2}=e^{\langle K\rangle/(2M_P)}|\langle W\rangle|/M_P^2
=e^{\sum_i|b_i|^2/2}m$.  

The soft SUSY breaking terms can read from 
the scalar potential in SUGRA: 
\dis{ \label{sugraPot}
V_F=e^{\frac{K}{M_P^2}}\left[\sum_i
\left|F_{z_i}\right|^2+\sum_r\left|F_{\phi_r}\right|^2
-\frac{3}{M_P^2}|W|^2\right] ,
}
where the ``$F$-terms,'' 
$F_X$ [$=(D_XW)^*=(\partial_XW+\partial_XK ~W/M_P^2)^*$] are,  
in the minimal SUGRA, given by 
\dis{
&F_{z_i}^*=\frac{\partial W_H}{\partial z_i}+z_i^*\frac{W}{M_P^2} 
=M_P\left[\left(a_i^*+b_i^*\right)m+b_i^*\frac{W_O}{M_P^2}\right] ,
\\  \label{F_a}
&F_{\phi_r}^*=\frac{\partial W_O}{\partial \phi_r}+\phi_r^*\frac{W}{M_P^2}
=\frac{\partial W_O}{\partial \phi_r}+\phi_r^*\left(m+\frac{W_O}{M_P^2}\right) . 
}
Note that VEVs of $F_{z_i}$ are of order ${\cal O}(mM_P)$. 
For the vanishing cosmological constant (C.C.), 
a fine-tuning between $\langle F_{z_i}\rangle$ and 
$\langle W_H\rangle$,  
$\sum_i\langle |F_{z_i}|^2\rangle=3|\langle W_H\rangle|^2/M_P^2$, or 
$\sum_{i}\left|a_i+b_i\right|^2=3$, 
is required from \eq{sugraPot}.
Neglecting the Planck-suppressed nonrenormalizable terms, 
\eq{sugraPot} is rewritten as \cite{PR1984} 
\dis{ \label{scalarPot}
V_F \approx 
\left|\partial_{\phi_r}\widetilde{W}_O\right|^2
+m_{0}^2|\phi_r|^2
+m_{0}\left[\phi_r\partial_{\phi_r}\widetilde{W}_O
+(A_\Sigma-3)\widetilde{W}_O+{\rm h.c.}\right] ,
}
where summations for $\phi_r$ are assumed.  
$A_\Sigma$ is defined as $A_\Sigma\equiv \sum_ib_i^*(a_i+b_i)$
and  
$m_0$ is identified with the gravitino mass $m_{3/2}$ 
($=e^{\sum_i|b_i|^2/2}m$).   
$\widetilde{W}_O$ ($\equiv e^{\sum_i|b_i|^2/2}W_O$) 
means the rescaled $W_O$. 
From now on, we will drop out the ``tilde'' for simplicity. 
In \eq{scalarPot}, the first term is nothing but  
the $F$-term scalar potential in global SUSY.  
The second and other terms imply that the 
soft scalar mass terms and soft SUSY breaking $A$-terms    
parametrized with $m_0$ 
are {\it universal} at the GUT scale in the mGrM. 
If there are no quadratic or higher powers of $\phi_r$ in $W_O$, 
one can get negative (positive) $A$-terms with $A_\Sigma <2$ ($A_\Sigma >2$). 
Here the {\it universal} $A$-parameter ($\equiv A_0=A_t$) 
does not include Yukawa coupling constants, 
but it is proportional to $m_0$.
We will set the universal $A$-term to 
\dis{
A_0\equiv a_Ym_0 ,
}  
where $a_Y$ is a dimensionless number. 
Using the vanishing C.C. condition, 
the universal soft mass parameter,  
$m_0$ ($=e^{\langle K\rangle/(2M_P^2)}\langle W_H\rangle/M_P^2$)
can be expressed as  
$e^{\langle K\rangle/(2M_P^2)}\left(\sum_i|\langle F_{z_i}\rangle|^2\right)^{1/2}/\sqrt{3}M_P$.  
It is the conventional form of $m_0$ in the mGrM scenario.

In $N=1$ SUGRA, the gauge kinetic function $f_{ab}$, 
which is a holomorphic function of scalar fields, 
not only determines the form of the gauge fields' 
kinetic terms 
[$=-\frac{1}{4}({\rm Re}f_{ab})F^{a\mu\nu}F^b_{\mu\nu}$], 
but also contributes to the gaugino mass term \cite{PR1984}: 
\dis{ \label{gauginoMass}
\frac{M_P}{4} ~e^{G/(2M_P^2)}
~\frac{\partial f^*_{ab}}{\partial z_i^*}
~\frac{\partial G}{\partial z_i} ~\lambda^a\lambda^b 
=\frac{1}{4}e^{\sum_i|b_i|^2/2}
~\frac{\partial f^*_{ab}}{\partial z_i^*} ~F_{z_i}^*
~\lambda^a\lambda^b ,
}
where $G$ is defined as $G\equiv K+M_P^2{\rm log}(W/M_P^3)$, 
and $\lambda^{a,b}$ stand for the gaugino fields. 
If SUSY is broken ($F_{z_i}\neq 0$) and 
the gauge kinetic function is nontrivial 
($\partial f_{ab}/\partial z_i\neq 0$),   
the gaugino masses can be generated. 
In the mGrM scenario, the unified gaugino mass $m_{1/2}$ 
is regarded as an independent parameter,     
assuming the canonical kinetic terms for the gauge fields. 
In our model that will be discussed in Section~\ref{sec:MMM},
however, we will employ the minimal form of the gauge kinetic function 
($=\delta_{ab}$) at tree level: 
the gaugino masses can be generated radiatively.  


\subsection{Focus Point in the Minimal Gravity Mediation}

As seen in \eq{scalarPot}, 
the soft SUSY breaking masses squared 
for the superpartners of chiral fermions are universal 
at the GUT scale in the mGrM.  
Accordingly, the $m_{q_30}^2$, $m_{u_3^c0}^2$, and $m_{h_u0}^2$ 
in \eqs{Sol1}, (\ref{Sol2}), and (\ref{Sol3}) 
should be set to be the same as $m_0^2$ in the mGrM: 
\dis{ \label{mGrMbdy}
m_{q_30}^2=m_{u^c_30}^2=m_{h_u0}^2=m_0^2 ,\quad {\rm and~~so} \quad X_0=3m_0^2 .
}
Thus, the semianalytic RG solutions take the following form: 
\dis{ \label{mhuGr}
&\quad~~ m_{h_u}^2(t)=\frac{3m_0^2}{2}\left[e^{\frac{3}{4\pi^2}\int^{t}_{t_0}dt^\prime y_t^2}-\frac13\right]
+\frac{F(t)}{2}
\\
&-\left(\frac{m_{1/2}}{g_0^2}\right)^2
\left[\frac32\bigg\{g_2^4(t)-g_0^4\bigg\}
+\frac{1}{22}\bigg\{g_1^4(t)-g_0^4\bigg\}\right]
}
and 
\bea \label{mSTopGr}
&&\qquad\quad~ \bigg\{m_{q_3}^2(t)+m_{u_3^c}^2(t)\bigg\}
=\frac{3m_0^2}{2}\left[e^{\frac{3}{4\pi^2}\int^{t}_{t_0}dt^\prime y_t^2}+\frac13\right]
+\frac{F(t)}{2}
\\
&&+ \left(\frac{m_{1/2}}{g_0^2}\right)^2\left[
\frac{16}{9}\bigg\{g_3^4(t)-g_0^4\bigg\}
-\frac32\bigg\{g_2^4(t)-g_0^4\bigg\}
-\frac{17}{198}\bigg\{g_1^4(t)-g_0^4\bigg\}\right] ,
\nonumber 
\eea
where $F(t)$ has been presented in \eq{F(t)}. 
The $A$-term contributions to the above solutions 
are all included in $F(t)$. 
The independent parameters in \eqs{mhuGr} and (\ref{mSTopGr}) 
are, thus, $m_0^2$, $(m_{1/2}/g_0^2)$, and $a_Y$: 
we regard $t_0$ (or $M_{G}$) as a given parameter, 
whose value is determined with the MSSM field contents 
and their interactions.  
Note that the above semianalytic solutions are valid 
only for small $\tb$ cases. 
For the solutions in larger $\tb$ cases, 
numerical analyses on the full RG equations should be implemented. 
Most of all, the above solutions are not valid any longer 
below the stop mass scale, 
since the stops should decouple from the RG equations: 
the RG equations should be modified below that scale.

In the original FP scenario \cite{FMM1}, 
it was pointed out that $e^{\frac{3}{4\pi^2}\int^{t}_{t_0}dt^\prime y_t^2}$ in \eq{mhuGr}
{\it happens to be almost $\frac13$} 
for $t\sim t_Z$ [$\equiv {\rm log}(M_Z/{\rm GeV})$], 
if the stops were not decoupled and \eq{mhuGr} was valid 
down to the $Z$ boson mass scale. 
In that case, the coefficient of $m_0^2$ in \eq{mhuGr} 
becomes very small, and so $m_{h_u}^2$ can almost be 
independent of $m_0^2$ around the $Z$ boson mass scale. 
It implies that a FP of $m_{h_u}^2(t)$ appears 
around the $Z$ boson mass scale. 
Note that the stop masses squared are quite sensitive 
to $m_0^2$ for $e^{\frac{3}{4\pi^2}\int^{t_Z}_{t_0}dt^\prime y_t^2}\approx\frac13$, 
as seen in \eq{mSTopGr}.
The coefficient of $(m_{1/2}/g_0^2)^2$ 
included in $F(t)/2$, 
which is generically bigger than those 
in the second line of \eq{mhuGr},  
turns out to be negative. 
Unlike the stop masses, therefore, 
$m_{h_u}^2$ can be naturally small at the $Z$ boson mass scale, 
only if $(m_{1/2}/g_0^2)$ and $a_Y$ are small enough.  

As mentioned in Introduction, however, 
the stop mass needs to be about $3$-$4 \tev$ 
for explaining the $126 \gev$ Higgs mass. 
It means that \eqs{mhuGr} and (\ref{mSTopGr}) are valid 
just down to $3$-$4 \tev$, and below the stop mass scale
the estimation \eq{RGsm} should be applied for $m_{h_u}^2$. 
This process would leave a sizable coefficient 
of $m_0^2$ in $m_{h_u}^2(t_Z)$, particularly in large $\tb$ cases. 
Hence a quite heavy stop mass would spoil 
the FP behavior of $m_{h_u}^2(t)$. 
To get a stop mass of $3$-$4 \tev$, moreover, 
$m_0^2$ needs to be large enough in \eq{mSTopGr}, 
which could require a large enough $(m_{1/2}/g_0^2)^2$ 
for EW symmetry breaking in large $\tb$ cases.

The coefficients of $m_0^2$, $(m_{1/2}/g_0^2)^2$, $\cdots$, etc. in \eqs{mhuGr} and (\ref{mSTopGr}) 
can numerically be calculated: 
\dis{ \label{NumGr}
&\qquad~~~ m_{h_u}^2(t_T)\approx \bigg[0.03 -0.11 a_Y^2\bigg]m_0^2
-0.25\left(\frac{m_{1/2}}{g_0^2}\right)^2
-0.16\left(\frac{m_{1/2}}{g_0^2}\right)a_Ym_0 ,
\\
&\left\{m_{q_3}^2(t_T)+m_{u_3^c}^2(t_T)\right\}
\approx \bigg[1.03 -0.11 a_Y^2\bigg]m_0^2
+1.20\left(\frac{m_{1/2}}{g_0^2}\right)^2
-0.16\left(\frac{m_{1/2}}{g_0^2}\right)a_Ym_0 ,
}
which are the values at the stop decoupling scale, 
$t=t_T\approx 8.2$ (i.e. $Q_T=3.5 \tev$) with $\tb=5$. 
From the above expression of $m_{h_u}^2(t_T)$, 
we can expect that a FP of $m_{h_u}^2$ appears 
below [above] $t_T$ (or $Q_T=3.5\tev$) 
when $a_Y^2<0.03/0.11\approx 0.27$ [$a_Y^2\gtrsim 0.27$].
As mentioned above, 
$\{m_{u_3^c}^2(t_T)+m_{q_3}^2(t_T)\}$ 
should be constrained to be around $2\cdot(3.5 \tev)^2$ 
in order to get the $126 \gev$ Higgs boson mass. 
While the stops masses would be frozen, thus, 
$m_{h_u}^2$ further decreases below the stop mass scale 
dominantly through the top quark Yukawa coupling: 
$m_{h_u}^2$ at the $Z$ boson mass scale can be estimated 
using \eq{RGsm}. 
It has the following structure:
\dis{ \label{STmGr}
m_{h_u}^2(t_Z)=C_s ~m_0^2
-C_g\left(\frac{m_{1/2}}{g_0^2}\right)^2
-C_m~a_Ym_0\left(\frac{m_{1/2}}{g_0^2}\right) ,
}
where the coefficients, $C_s$, $C_g$, and $C_m$ 
are approximately given by  
\dis{
&\qquad~ C_s\approx 0.03-0.11a_Y^2-\frac{3|y_t|^2}{16\pi^2}\times\left(
1.03-0.11 a_Y^2\right) ,
\\
&C_g\approx 0.25 +\frac{3|y_t|^2}{16\pi^2}\times 1.20 ,
\quad {\rm and} \quad 
C_m\approx 0.16 -\frac{3|y_t|^2}{16\pi^2}\times 0.16 ,
}
for $\tb=5$. 
Since the SU(3)$_c$ gauge coupling becomes almost unity 
around the $3.5 \tev$ energy scale, 
$(m_{1/2}/g_0^2)$ in the above equations can approximately 
be regarded as the low energy gluino (running) mass:  
\dis{
\frac{m_{1/2}}{g_0^2}=\frac{M_3(t_T)}{g_3^2(t_T)}\approx M_3(t_T) .
}
For $m_0^2\gg M_3^2(t_T)$ and $a_Y^2\ll 1$,  
$m_0^2\sim (4.2 \tev)^2$--$(5.6 \tev)^2$ is needed 
for $3$-$4 \tev$ stop masses in \eq{NumGr}.
%
Although the semianalytic solutions, 
\eqs{mhuGr} and (\ref{mSTopGr})
are not valid any longer for large $\tb$ cases, 
the basic structure of $m_{h_u}^2(t_Z)$ in those cases 
would still have the form of \eq{STmGr}, 
but with different values for $C_s$, $C_g$, and $C_m$ 
from \eq{STmGr}.

%
%
\begin{figure}
\begin{center}
\includegraphics[width=0.6\linewidth]{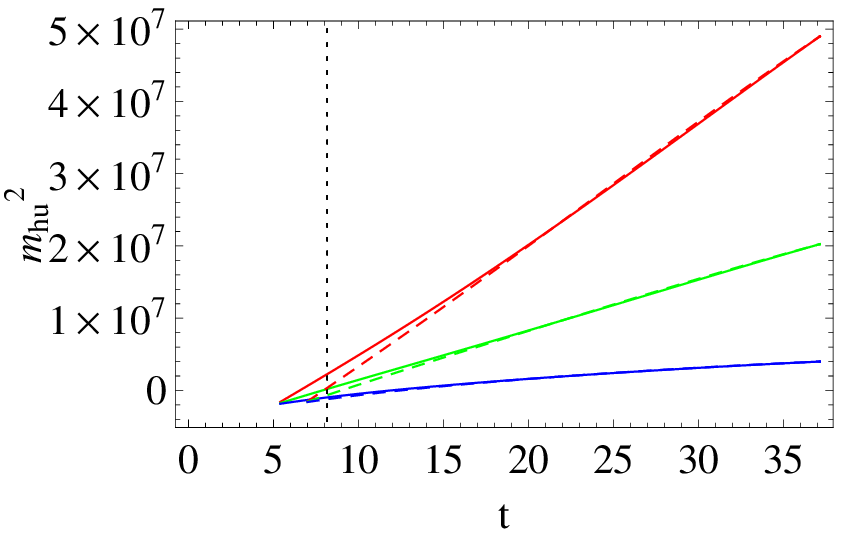}
\end{center}
\caption{RG evolutions of $m_{h_u}^2$ in the mGrM 
with $t$ [$\equiv {\rm log}(Q/\gev)]$
for $m_0^2=(7 \tev)^2$ [red], $(4.5 \tev)^2$ [green], 
and $(2 \tev)^2$ [blue], 
when $m_{1/2}/g_0^2=2.3 \tev$ and $A_t=0$ at the GUT scale. 
The tilted solid [dotted] lines correspond to the case of 
$\tb=50$ [$\tb=5$]. 
The vertical dotted line at $t=t_T\approx 8.2$ 
($Q_T=3.5 \tev$) indicates the desired stop mass scale. 
Below the stop decoupling scale, in fact, the RG evolutions 
should be modified from this figure. 
The FP of $m_{h_u}^2$ would appear 
around $t\approx 5.3$ ($Q\approx 200 \gev$) 
[$t\approx 7.0$ ($Q\approx 1.1 \tev$)], however,  
if its RG evolutions are extrapolated below $t=t_T$, 
keeping heavy superpartners. 
}
\label{fig:mGrM}
\end{figure}
%
%

Figure~\ref{fig:mGrM} displays the full numerical results 
on the RG behaviors of $m_{h_u}^2(t)$ for $\tb=50$ (solid lines) and $\tb=5$ (dotted lines) 
under various trial $m_0^2$, based on the full RG equations 
including $y_{b,\tau}$, $A_{b,\tau}$, $m_{b,\tau,h_d}^2$, etc., 
when $(m_{1/2}/g_0^2)=2.3 \tev$ and $A_t=a_Y=0$ at the GUT scale. 
In fact the RG runnings of of $m_{h_u}^2(t)$ had to be 
modified below the stop decoupling scale. 
Nonetheless, we extrapolate $m_{h_u}^2(t)$s below $t=t_T$, 
keeping heavy superpartners in the RG evolutions,  
in order to discuss the FPs of $m_{h_u}^2$.  
As seen in Fig.~\ref{fig:mGrM}, the FP appears at a scale relatively close to $t_T$ for $\tb=5$, when $a_Y=0$. 
That is the reason why the coefficient of $m_0^2$ 
in $m_{h_u}^2(t_T)$ of \eq{NumGr} is small. 
For $\tb=50$, thus, we can expect that 
the coefficient of $m_0^2$ is quite sizable, 
since the FP is relatively far from $t_T$.    
 
From \eq{STmGr}, we see that
the gluino mass should be heavier than $1.3 \tev$ 
for EW symmetry breaking, i.e. $m_{h_u}^2(t_Z)<0$ 
with $m_0^2\sim (4.5 \tev)^2$ and $a_Y^2\ll 1$.  
To meet the experimental bound $M_3(t_T)>1.3 \tev$, 
therefore, $\tb$ should be larger than 5,  
when stop masses are $3$-$4 \tev$ stop masses and $|a_Y|\ll 1$.
For larger $\tb$ cases, heavier low energy gluino masses  
are necessary for EW symmetry breaking. 
Since $y_{b,\tau}$, $A_{b,\tau}$, etc. are quite small in small $\tb$ cases, however,  
the RG evolution of $m_{h_d}^2$ would be negligible 
and so its low energy values are almost the same as $m_0^2$.  
As a result, $|\mu|$ consistent with $m_Z\approx 91 \gev$ 
in \eq{m_Z} exceeds $900 \gev$ 
for $\tb=5$ and $m_0^2=(4.5 \tev)^2$. 
A larger $m_0^2$ or a larger $(m_{1/2}/g_0^2)^2$ 
requires a larger $|\mu|^2$ in general. 

In fact, the RG equation of $\mu$ is completely separated from those of the soft parameters at one-loop level. 
Moreover, its generation scale is quite model dependent. 
Thus, we do not discuss them in this paper. 
To avoid a potentially problematic fine-tuning issue 
associated with $\mu$, however, 
we will consider only the cases of 
$\frac12 m_Z^2/|\mu|^2>0.01$ or $|\mu|<600 \gev$. 
Numerical analyses show that $\tb$ should be larger than $8$ 
for $|\mu|<600 \gev$, when $m_0^2=(4.5 \tev)^2$ 
and $|a_Y|\ll 1$. 
In this case, the low energy gluino mass should be heavier 
than $1.9 \tev$ for EW symmetry breaking.

 
%
%
 
Since the coefficients of $m_0^2$ change slowly 
under a small variation $\delta m_0^2$,  
the small change of $\delta m_{h_u}^2$ under $\delta m_0^2$ 
at the $Z$ boson mass scale is roughly estimated as 
\dis{ \label{FPmGrM}
\frac{\delta m_{h_u}^2}{\delta m_0^2} \approx
C_s-\frac{a_YC_m}{2m_0}\left(\frac{m_{1/2}}{g_0^2}\right) ,
}
which makes contribution to the fine-tuning measure \cite{FTmeasure}, 
\dis{ \label{Deltam02}
\Delta_{m_0^2}=\frac{\delta ~{\rm log}m_Z^2}{\delta ~{\rm log}m_0^2}
=\frac{m_0^2}{m_Z^2}~\frac{\delta m_Z^2}{\delta m_0^2}
=2\left(\frac{m_0^2}{m_Z^2}\right)\left[
\frac{(\delta m_{h_d}^2/\delta m_0^2)-\tan^2\beta(\delta m_{h_u}^2/\delta m_0^2)}{\tan^2\beta-1}\right] . 
}
Note that $(m_0^2/m_Z^2)$ is a very large number,
because a quite large $m_0^2$ [$(4.2 \tev)^2$-$(5.6 \tev)^2$] 
is necessary for a $3$-$4 \tev$ stop mass.
Hence, the other parts in \eq{Deltam02} should sufficiently be 
suppressed to get a small enough $\Delta_{m_0^2}$.
As clearly seen in \eq{FPmGrM}, the variation of $m_{h_u}^2$ 
under $\delta m_0^2$, $(\delta m_{h_u}^2/\delta m_0^2)$ 
cannot be zero at the stop mass scale, 
unless $a_Y$ is finely tuned. 
As mentioned above, moreover, low energy values of $m_{h_d}^2$  
are almost the same as $m_0^2$s in small $\tb$ cases. 
Accordingly, $(\delta m_{h_d}^2/\delta m_0^2)$ 
would be about unity in \eq{Deltam02}.
Therefore, {\it $\Delta_{m_0^2}$ and $|\mu|$ 
cannot be small enough in small $\tb$ cases, 
when stop masses are $3$-$4 \tev$ or heavier}. 

In large $\tb$ cases, $(\delta m_{h_d}^2/\delta m_0^2)$ 
is relatively suppressed as seen in \eq{Deltam02}. 
In fact, $m_{h_d}^2$ is not focused at all. 
Hence, a larger $\tb$ would be more desirable 
in the FP scenario. 
In the case of $\tb=50$, for instance, 
the physical [low energy running] gluino mass should be 
heavier than $2.6 \tev$ [$2.2 \tev$] 
for EW symmetry breaking, 
but lighter than $2.8 \tev$ [$2.6 \tev$] for $|\mu|<600 \gev$, 
when $m_0^2=(4.5 \tev)^2$ and $|a_Y|\ll 1$.
%
However, 
the FP scale is basically too far from the stop mass scale 
as shown in Fig.~\ref{fig:mGrM}. 
Consequently, $(\delta m_{h_u}^2/\delta m_0^2)$ 
in \eq{Deltam02} or $C_s$ in \eq{STmGr} is quite sizable,     
and so $\Delta_{m_0^2}$ is hard to be small enough 
also in large $\tb$ cases. 
We should note here that a sizable $C_s$ in \eq{STmGr}   requires also a sizable $C_g(m_{1/2}/g_0^2)^2$ or 
$C_ma_Y(m_{1/2}/g_0^2)$ for EW symmetry breaking.

\begin{table}[!h]
\begin{center}
\begin{tabular}
{c|ccc}
\hline\hline
{\footnotesize ${\bf A_0/m_0=0.3}$} & {\footnotesize $M_3(t_T)=2.5 \tev$}~  & {\footnotesize ${\bf |\mu|=903 \gev}$}  & {\footnotesize ${\bf \Delta_{m_0^2}=276}$}  
\\ 
 {\footnotesize ${\bf m_0^2}$} & {\footnotesize $({\bf 5.5} \tev)^2$}  & {\footnotesize $({\bf 4.5} \tev)^2$}  & {\footnotesize $({\bf 3.5} \tev)^2$}  
\\ \hline
 {\footnotesize $m_{q_3}^2(t_T)$} &  {\footnotesize $(4437 \gev)^2$}  & {\footnotesize $(3817 \gev)^2$}  & {\footnotesize $(3238 \gev)^2$} 
\\
 {\footnotesize $m_{u^c_3}^2(t_T)$} &  {\footnotesize $(3857 \gev)^2$}  & {\footnotesize $(3329 \gev)^2$}  & {\footnotesize $(2839 \gev)^2$} 
\\
 {\footnotesize ${\bf m_{h_u}^2(t_T)}$} & {\footnotesize $({\bf 461} \gev)^2$}  & {\footnotesize $-({\bf 694} \gev)^2$}  & {\footnotesize $-({\bf 1007} \gev)^2$}  
\\
 {\footnotesize $m_{h_d}^2(t_T)$} &  {\footnotesize $(2585 \gev)^2$}  & {\footnotesize $(2032 \gev)^2$}  & {\footnotesize $(1450 \gev)^2$}
\\ \hline\hline
{\footnotesize ${\bf A_0/m_0=0}$} & {\footnotesize $M_3(t_T)=2.5 \tev$}~  & {\footnotesize ${\bf |\mu|=387 \gev}$}  & {\footnotesize ${\bf \Delta_{m_0^2}=378}$}  
\\ 
 {\footnotesize ${\bf m_0^2}$} & {\footnotesize $({\bf 5.5} \tev)^2$}  & {\footnotesize $({\bf 4.5} \tev)^2$}  & {\footnotesize $({\bf 3.5} \tev)^2$}  
\\ \hline
 {\footnotesize $m_{q_3}^2(t_T)$} &  {\footnotesize $(4497 \gev)^2$}  & {\footnotesize $(3870 \gev)^2$}  & {\footnotesize $(3285 \gev)^2$} 
\\
 {\footnotesize $m_{u^c_3}^2(t_T)$} &  {\footnotesize $(3933 \gev)^2$}  & {\footnotesize $(3396 \gev)^2$}  & {\footnotesize $(2897 \gev)^2$} 
\\
 {\footnotesize ${\bf m_{h_u}^2(t_T)}$} & {\footnotesize $({\bf 1044} \gev)^2$}  & {\footnotesize $({\bf 442} \gev)^2$}  & {\footnotesize $-({\bf 721} \gev)^2$}  
\\
 {\footnotesize $m_{h_d}^2(t_T)$} &  {\footnotesize $(2749 \gev)^2$}  & {\footnotesize $(2189 \gev)^2$}  & {\footnotesize $(1607 \gev)^2$}
\\ \hline\hline
{\footnotesize ${\bf A_0/m_0=-1.0}$} & {\footnotesize $M_3(t_T)=2.5 \tev$}~  & {\footnotesize ${\bf |\mu|=753 \gev}$}  & {\footnotesize ${\bf \Delta_{m_0^2}=83}$}  
\\ 
 {\footnotesize ${\bf m_0^2}$} & {\footnotesize $({\bf 5.5} \tev)^2$}  & {\footnotesize $({\bf 4.5} \tev)^2$}  & {\footnotesize $({\bf 3.5} \tev)^2$}  
\\ \hline
 {\footnotesize $m_{q_3}^2(t_T)$} &  {\footnotesize $(4427 \gev)^2$}  & {\footnotesize $(3840 \gev)^2$}  & {\footnotesize $(3289 \gev)^2$} 
\\
 {\footnotesize $m_{u^c_3}^2(t_T)$} &  {\footnotesize $(3840 \gev)^2$}  & {\footnotesize $(3354 \gev)^2$}  & {\footnotesize $(2900 \gev)^2$} 
\\
 {\footnotesize ${\bf m_{h_u}^2(t_T)}$} & {\footnotesize $({\bf 105} \gev)^2$}  & {\footnotesize $-({\bf 478} \gev)^2$}  & {\footnotesize $-({\bf 702} \gev)^2$}  
\\
 {\footnotesize $m_{h_d}^2(t_T)$} &  {\footnotesize $(2385 \gev)^2$}  & {\footnotesize $(1952 \gev)^2$}  & {\footnotesize $(1498 \gev)^2$}
\\ \hline\hline
%
\end{tabular}
\end{center}\caption{Soft squared masses of the stops and Higgs bosons at $t=t_T\approx 8.2$ ($Q_T=3.5 \tev$) in the mGrM 
for various trial $m_0^2$s when $\tb=50$.  
$\Delta_{m_0^2}$ indicates the fine-tuning measure for $m_0^2$ around ${(4.5 \tev)^2}$ 
for each case. 
%
}\label{tab:mGrM2}
\end{table}
%



Table~\ref{tab:mGrM2} lists soft squared masses of the stops and Higgs bosons at $t=t_T\approx 8.2$ ($Q_T=3.5 \tev$) 
for various trial $m_0^2$s and $A_0$, 
when $\tb=50$ and $M_3(t_T)=2.5 \tev$. 
They are results generated by {\tt SOFTSUSY}-3.6.2 \cite{softsusy}, 
analyzing the full RG equations. 
We can see that 
$\Delta_{m_0^2}$s for $m_{h_u}^2$ 
are of order $10^2$ for $|\mu|<600 \gev$. 
It is because the FP of $m_{h_u}^2$ appears 
too far below $t=t_T$ as discussed above.
%

To summarize, $|\mu|$ and $\Delta_{m_0^2}$  
are too large in small $\tb$ cases in the mGrM, 
even if the FP emerges somewhat close to the stop mass scale. 
It is because the $m_0^2$ needed for the desired stop mass 
is quite heavy, 
and $m_{h_d}^2$ ($\approx m_0^2$) is not focused at all. 
In large $\tb$ cases, on the other hand, 
the FP scale of $m_{h_u}^2$ is too low, 
compared with the stop mass scale. 

To keep a small enough $\mu$ even with $3$-$4 \tev$ stop masses, 
thus, we should consider a large $\tb$ case. 
But we need to somehow push the FP scale up to the desired 
stop mass scale in order to reduce $\Delta_{m_0^2}$ in this case. 
Of course, there still remains a possibility to achieve it 
by assuming a (fine-tuned) $a_Y$ with a large $\tb$. 
A fine-tuned Dirac Yukawa coupling of a RH neutrino, $y_N$ 
is also helpful for pushing the FP \cite{RHneu,KS}. 
However, it is very hard to contrive a model 
to naturally explain such a special value of $a_Y$ or $y_N$,  
reducing also $\Delta_{A_0}$ or $\Delta_{y_N}$.  
In the next section, we will propose another way 
to move the FP scale up to the desired stop mass scale 
in a large $\tb$ case.

\section{Minimal Mixed Mediation} \label{sec:MMM}

In large $\tb$ cases, as mentioned above, 
$C_s$ is sizable in \eq{STmGr} 
because the FP of $m_{h_u}^2$ is far below the stop decoupling scale, 
and $C_g(m_{1/2}/g_0^2)^2$ and/or $C_ma_Y(m_{1/2}/g_0^2)$ 
are also required to be large enough 
for EW symmetry breaking. 
While the $C_s$ term makes a positive contribution to 
$m_{h_u}^2(t_Z)$ for small $a_Y$s, 
the other terms make negative contributions to it. 
In this section, we will attempt to investigate  
a mechanism in which the two sizable contributions can  
automatically be canceled 
to eventually yield a small enough $C_s$ 
even in a large $\tb$ case.

\subsection{Basic Setup in the Minimal Mixed Mediation}

On top of the mGrM setup, we consider also the mGgM effects 
by introducing one pair of messenger fields 
$\{{\bf 5}_M, \overline{\bf 5}_M\}$
which are the SU(5) fundamental representations, 
Through their coupling with an MSSM singlet superfield $S$,   
\dis{ \label{Wm}
W_m=y_SS{\bf 5}_M\overline{\bf 5}_M ,
}
the soft masses of the MSSM gauginos 
and scalar superpartners are generated 
at one- and two-loop levels, respectively, 
if the scalar and $F$-term components of $S$ develop 
nonzero VEVs \cite{book}:
\dis{ \label{GGsoftmass}
M_a|_M=\frac{g_a^2(t_M) ~\langle F_S\rangle}{16\pi^2\langle S\rangle} ~, 
\qquad 
\delta m_{\phi_r}^2|_M=2\sum_{a=1}^3 \left[\frac{g_a^2(t_M) ~\langle F_S\rangle}{16\pi^2\langle S\rangle}\right]^2 C_a(r)
}
where $C_a(r)$ denotes the quadratic Casimir invariant for a superfield $\Phi_r$, $(T^aT^a)^{r^\prime}_r=C_a(r)\delta^{r^\prime}_r$, and $g_a$ ($a=3,2,1$) is the MSSM gauge couplings. 
$\langle S\rangle$ and $\langle F_S\rangle$ are VEVs of the scalar and $F$-term components of the superfield $S$. 
Note that $M_a$ and $m_{\Phi_r}^2$ are almost independent of $y_S$ 
only if $\langle F_S\rangle\lesssim y_S\langle S\rangle^2$ \cite{book}. 
However, such mGgM effects appear 
below the messenger mass scale, $y_S\langle S\rangle$.  
In this paper, we assume the messenger mass scale is 
lower than the GUT scale. 
Otherwise, $\delta m_{\phi_r}^2|_M$ as well as $M_a|_M$ could 
become relatively universal at the GUT scale (as in the mGrM), 
respecting the relations required by a given GUT,  
since non-MSSM gauge sectors contained in a SUSY GUT 
such as ``$X$'' and ``$Y$'' in the SU(5) GUT also contribute to 
$\delta m_{\phi_r}^2|_M$.

Once the hidden sector superpotential $W_H$ develops a VEV, 
the $F$-term of $S$ as well as the $F$-terms of superfields 
in the hidden sector can also get VEVs proportional 
to $\langle W_H\rangle$ ($\equiv mM_P^2$). 
For instance, let us consider the following  K${\rm\ddot{a}}$hler potential in addition to \eq{KSpot}: 
\dis{ \label{Extpot}
K\supset f(z)S +{\rm h.c.} ,
}      
where $f(z)$ is a {\it holomorphic} monomial of hidden sector fields $z_i$s 
with VEVs of order $M_P$ in \eq{vev}, 
and so $f(z)$ should be of order ${\cal O}(M_P)$.  
Its specific form can be controlled 
by introducing hidden local symmetries.
Note that the above term leaves intact the kinetic terms of $z_i$s, and so they still remain as the canonical form. 
$M_Pf(z)S$ in the superpotential can be forbidden 
by the U(1)$_R$ symmetry.
By including the SUGRA corrections with $\langle W_H\rangle=mM_P^2$, 
then, $\langle F_S\rangle$ can be  
\dis{
\langle F_S^*\rangle\approx m\left[\langle f(z)\rangle+\langle S^*\rangle\right] ,
}
if $\langle\partial_SW\rangle$ is relatively suppressed 
by relevant small (or zero) Yukawa couplings.
Thus, the VEV of $F_S$ is of order ${\cal O}(mM_P)$ like $F_{z_i}$ in \eq{F_a}. 
They should be fine-tuned for the vanishing C.C.:  
a precise determination of $\langle F_S\rangle$ is indeed 
associated with the C.C. problem.
%
%
Here we set $\langle F_S\rangle=m_0M_P$. 
%
$F_{\phi_r}$ is still given by \eq{F_a}, 
which induces the universal soft mass terms at tree level 
for the observable scalar fields. 
Consequently, both the gravity and gauge mediation effects 
are induced from a {\it single} SUSY breaking source, 
and they all are parametrized with $m_0$.  

%
%
%
\begin{table}[!h]
\begin{center}
\begin{tabular}
{c|ccccc|cccc|c}
  & $S$ & $S^c$ & ${\bf 24}^\prime$ & ${\bf 24}$ & ${\bf 24}^c$ 
  & $z$ & $\bar{z}$ & $z^c$ & $\bar{z}^{c}$ & ~$\Sigma_R$    
\\ \hline 
U(1)$_Z$ & $+1$ & $-1$ & $0$ & $+1$ & $-1$ & $\frac{+1}{2}$ & $\frac{+1}{2}$ & $\frac{-1}{2}$ & $\frac{-1}{2}$ & $0$    
\\
G$_H$  & ${\bf 1}$ & ${\bf 1}$ & ${\bf 1}$ & ${\bf 1}$ & ${\bf 1}$ & 
$R$ & $\overline{R}$ & $R$ & $\overline{R}$ & ${\bf 1}$   
\\
U(1)$_R$  & $0$ & $1$ & $2$ & $1$ & $0$ & $0$ & $0$ & $0$ & $0$ & $-2$    

\end{tabular}
\end{center}\caption{Quantum numbers of superfields for a local U(1)$_Z$, a hidden gauge G$_H$, and the global U(1)$_R$ symmetries. 
Only the hidden sector fields  
$\{z,\bar{z},z^c,\bar{z}^c\}$ 
carry proper nontrivial quantum numbers $\{R,\overline{R}\}$ 
under the hidden gauge group G$_H$.
}\label{tab:charges}
\end{table}
%
%

We assume that $\langle S\rangle$ has the same magnitude as
the VEV of the SU(5) breaking Higgs ($\equiv v_G$),  
$\langle {\bf 24}_H\rangle=v_G\times {\rm diag.}(2,2,2;-3,-3)/\sqrt{60}$. 
It can be realized by constructing a proper model, 
in which a GUT breaking mechanism causes $\langle S\rangle$. 
%
%
For example, let us consider the following 
K${\rm\ddot{a}}$hler potential and superpotential: 
\bea \label{GUTbreakingW}
K&\supset& 
z^c\bar{z}^cS +{\rm h.c.} ,  
\nonumber \\
W&\supset& (z\bar{z})^2S^cS^c + (z^c\bar{z}^c)^2{\rm Tr}\left[{\bf 24}{\bf 24}\right] 
+\Sigma_R{\rm Tr}\left[{\bf 24}^\prime{\bf 24}^\prime\right]
\\
&& +{\rm Tr}\left[{\bf 24}^\prime\left\{\left(S+\lambda z\bar{z}\right){\bf 24}^c
-(z\bar{z})^2{\bf 24}^c{\bf 24}^c\right\}\right] ,
\nonumber
\eea
where we drop the ${\cal O}(1)$ dimensionless coupling constants 
and set $M_P=1$
for simple expressions except for 
$\lambda$ [$\sim 10^{-2}$]. 
Here we introduced a U(1)$_Z$ gauge symmetry 
and supposed that {\it some} hidden sector fields  
$\{z,\bar{z},z^c,\bar{z}^c\}$ [$\subset \{z_i\}$ in \eq{KSpot}], 
which are nontrivial representations of a hidden gauge group G$_H$ 
($\{R,\overline{R}\}$), 
carry U(1)$_Z$ charges as well. 
We also introduce the global U(1)$_R$ symmetry as well as the SU(5) visible gauge symmetry \cite{SO(10)}, 
under which $\{z,\bar{z},z^c,\bar{z}^c\}$ remain neutral. 
The other relevant superfields and their charges 
are presented in Table~\ref{tab:charges}.   
$\{{\bf 24}^\prime,{\bf 24},{\bf 24}^c\}$ are all 
SU(5) adjoint representations, while $\{S,S^c\}$ are singlets. 
$\Sigma_R$ denotes a spurion field, whose VEV  
breaks the U(1)$_R$ to the $Z_2$ symmetry.  
$W_m$ in \eq{Wm} can be reproduced   
by assigning the unit U(1)$_R$ charge 
to $\{{\bf 5}_M, \overline{\bf 5}_M\}$  
from $W_m=z^c\bar{z}^cS{\bf 5}_M\overline{\bf 5}_M$. 
Note that the field contents in Table~\ref{tab:charges} do not yield any gauge anomaly.

As in $\{z_i\}$ of \eq{vev}, $\{z,\bar{z},z^c,\bar{z}^c\}$ 
in \eq{GUTbreakingW} are assumed to get VEVs of the Planck scale.   
Note that the combinations of them, $z\bar{z}^c$ ($\equiv u$) and 
$\bar{z}z^c$ ($\equiv v$) do not carry any quantum numbers. 
Thus, the K${\rm\ddot{a}}$hler potential and superpotential 
in the hidden sector would take the forms of  $K_H=K_H(u,v)$ 
and $W_H=W_H(u,v)$, neglecting the asymmetric term 
$K\supset z^c\bar{z}^cS +{\rm h.c.}$ 
because of its smallness: 
the consistency of $\langle S\rangle\ll M_P$ will be confirmed.  
Accordingly, the $F$-terms of $\{z,\bar{z},z^c,\bar{z}^c\}$ are 
given by $F_z^*=\partial_zW_H+W_H\partial_zK_H
=\bar{z}^c(\partial_uW_H+W_H\partial_uK_H)$, 
$F_{\bar{z}^c}^*=\partial_{\bar{z}^c}W_H+W_H\partial_{\bar{z}^c}K_H
=z(\partial_uW_H+W_H\partial_uK_H)$, etc., 
which are all assumed to be of order ${\cal O}(mM_P)$. 
Since $|z|=|\bar{z}^c|$ minimizes $|F_z|^2+|F_{\bar{z}^c}|^2$ 
[$=(|z|^2+|\bar{z}^c|^2)|\partial_uW_H+W_H\partial_uK_H|^2$], 
$\langle z\rangle$ and $\langle\bar{z}^c\rangle$ would be developed 
along the direction of $|\langle z\rangle|=|\langle\bar{z}^c\rangle|$.
Note that the minimization of $|\partial_uW_H+W_H\partial_uK_H|^2$ 
would determine just $u$ or $v$.  
Similarly, $\langle z^c\rangle$ and 
$\langle\bar{z}^c\rangle$ would be developed 
along the $|\langle z^c\rangle|=|\langle\bar{z}^c\rangle|$ direction,  
minimizing $|F_{\bar{z}}|^2+|F_{z^c}|^2$. 
Moreover, such directions are the $D$-flat directions of G$_H$. 
Although the full $F$-term potential could be further minimized, 
both $|\langle z\rangle|=|\langle\bar{z}^c\rangle|$ and 
$|\langle z^c\rangle|=|\langle\bar{z}^c\rangle|$ 
should still be maintained. 
%

Due to the mass terms by the VEVs of $\{z,\bar{z},z^c,\bar{z}^c\}$ 
and $\Sigma_R$ in the superpotential of \eq{GUTbreakingW}, 
then, we have $\langle S^c\rangle=\langle{\bf 24}\rangle
=\langle{\bf 24}^\prime\rangle=0$ 
even after including the SUGRA corrections. 
On the other hand, ${\bf 24}^c$ can develop a VEV of the order GUT scale 
in the U(1)$_Y$ direction 
from the second line of $W$ in \eq{Extpot} 
as in the ordinary minimal SU(5) GUT \cite{GUT}.  
It is identified with ${\bf 24}_H$ discussed above. 
%
Both $\langle {\bf 24}^c\rangle$ and $\langle S\rangle$ are 
completely determined by the minimum conditions for $F_{{\bf 24}^\prime}$ and the $D$-term of U(1)$_z$ \cite{PR1984},  
\dis{ \label{Dterm}
D_z=g_z\sum_jq_j\left[\partial_{\varphi_j}K+M_P^2\frac{\partial_{\varphi_j}W}{W}\right]\varphi_j
=g_z\left(|S|^2-{\rm Tr}|{\bf 24}^c|^2 + \cdots \right) ,
}
where $g_z$ and $q_j$ mean the U(1)$_z$ gauge coupling and charge of a field $\varphi_j$.
``$\cdots$'' contains the contributions by $\{z,\bar{z},z^c,\bar{z}^c\}$ 
and other scalar fields with zero VEVs.
However, the VEVs of $z_i$ are canceled out from \eq{Dterm} 
because of $|\langle z\rangle|=|\langle\bar{z}^c\rangle|$ and 
$|\langle z^c\rangle|=|\langle\bar{z}^c\rangle|$.  
In the SUSY limit, thus, all the VEVs of the fields in Table~\ref{tab:charges} 
have been determined: $\langle S\rangle=v_G$ and others are vanishing.  
By including the SUGRA corrections by $\langle W_H\rangle=mM_P^2$, 
we can read the SUSY breaking effects:
$\langle F_S^*\rangle
=m\left(
z^c\bar{z}^c+S^*\right) \gg  
\langle F_{{\bf 24}^c}\rangle =m v_G$.  
Thus, VEVs of $F_S$ and $\{F_z,F_{\bar{z}},F_{z^c},F_{\bar{z}^c}\}$ are all ${\cal O}(mM_P)$. 
They should be fine-tuned for the vanishing C.C.:  
precise determination of $\langle F_S\rangle$ 
is associated with the C.C. problem as mentioned above.

$v_G$ induces the superheavy masses of $X$ and $Y$ gauge bosons and their superpartners in the SU(5) GUT, $M_X$ and $M_Y$.
%
Since the GUT gauge interactions would become active 
above their mass scale,  
$M_X^2=M_Y^2=\frac{5}{24}g_G^2v_G^2$ \cite{GUT}, 
it is identified with the MSSM gauge coupling unification scale. 
Thus, $\langle S\rangle$ ($=v_G$) is {\it fixed} 
by the relation with the unification scale.
When the superpartners of the SM chiral fermions are heavier than $3$-$4 \tev$, 
the unification scale is about $(0.9$-$1.7)\times 10^{16} \gev$. 
In fact, the three MSSM gauge couplings are not 
exactly unified at a unique scale 
only with the MSSM field contents, 
because the superpartners are relatively heavy in this case. 
However, various threshold effects would arise   
around that scale. 
Here we will take the central value of the above range, 
i.e.  $1.3\times 10^{16} \gev$ for the unification scale.  
Then the mGgM SUSY breaking effects in \eq{GGsoftmass} 
can be estimated with a parameter $f_G$:   
\dis{ \label{GGscale}
f_G \cdot m_0\equiv
\frac{\langle F_S\rangle}{16\pi^2\langle S\rangle}
=\frac{m_0 M_P}{16\pi^2M_X}\sqrt{\frac{5}{24}}g_{G}
\approx 0.36 ~m_0 .
}
Note that the $m_0$ dependence appears 
because $F_S$ is proportional to $m_0$ in the Minimal Mixed Mediation as discussed above. 
$f_G$ is basically a parameter determined by a model. 
From now on, however, we will leave $f_G$ as an unknown parameter.
%
%

From \eq{GGsoftmass}, 
the soft squared masses for the MSSM Higgs 
and the superpartners of (the third generation of) chiral fermions at the messenger scale are expressed as follows:      
\begin{eqnarray} 
&&\delta m_{h_u}^2|_{M}=\delta m_{h_d}^2|_{M}
=\delta m_{l_3}^2|_{M}=
f_G^2m_0^2 
\left[\frac{3}{2}g_2^4(t_M)+\frac{3}{10}g_1^4(t_M)\right] ,
\label{mGgM3} \\
&&\delta m_{q_3}^2|_{M}=
f_G^2m_0^2 
\left[\frac{8}{3}g_3^4(t_M)+\frac{3}{2}g_2^4(t_M)+\frac{1}{30}g_1^4(t_M)\right] ,
\label{mGgM1} \\
&&\delta m_{u_3^c}^2|_{M}=
f_G^2m_0^2 
\left[\frac{8}{3}g_3^4(t_M)+\frac{8}{15}g_1^4(t_M)\right] ,
\label{mGgM2} \\
&&\delta m_{d_3^c}^2|_{M}=
f_G^2m_0^2 
\left[\frac{8}{3}g_3^4(t_M)+\frac{2}{15}g_1^4(t_M)\right] ,
\\
&&\delta m_{e_3^c}^2|_{M}=
f_G^2m_0^2 
\left[\frac{6}{5}g_1^4(t_M)\right] ,
\label{mGgMe^c}
\end{eqnarray}
where $g_a(t_M)$s ($a=3,2,1$) denote the MSSM gauge coupling 
constants at the messenger scale. 
Hence, $\delta X|_M$ 
($\equiv \delta m_{q_3}^2|_{M}+\delta m_{u_3^c}^2|_{M}+\delta m_{h_u}^2|_{M}$) is given by 
\dis{
\delta X|_M
=f_G^2m_0^2\left[\frac{16}{3}g_3^4(t_M)+3g_2^4(t_M)+\frac{13}{15}g_1^4(t_M)\right] .
}
Note that the above soft masses, \eqs{mGgM3}-(\ref{mGgMe^c}) are not universal even around the GUT scale unlike the mGrM, 
since only the MSSM gauge sector makes contributions to $\delta m_{\phi_r}^2|_M$ and superheavy gauge sectors contained in a SUSY GUT 
would decouple at the GUT scale.

In contrast to the soft masses for the superpartners of SM chiral fermions, 
the gaugino masses are assumed to be generated dominantly 
only by the mGgM effect, i.e., $M_a$ of \eq{GGsoftmass}.
It is possible by employing the constant gauge kinetic function
($=\delta_{ab}$) at tree level, 
which is the {\it minimal} gauge kinetic function, 
yielding the canonical kinetic terms for gauge fields. 
Above the messenger mass scale, hence, 
the gaugino mass contributions to the RG equation 
should be negligible: the gaugino masses 
via mGrM must be small as seen in \eq{gauginoMass}.   
On the contrary, $A$-terms in the mGgM  
are generically much suppressed compared to those in the mGrM \cite{book}. 
So the universal $A$-terms coming from \eq{scalarPot}, 
which are proportional to $m_0$, should be dominant ones. 

Since the MSSM RG equations are valid below the messenger 
scale, the boundary conditions at the messenger scale,  
Eqs.~(\ref{GGsoftmass}) and (\ref{GGscale}) yield 
\dis{
\frac{M_a(t)}{g_a^2(t)}=\frac{M_a(t_M)}{g_a^2(t_M)}
=f_G \cdot m_0 .
}
Hence, the low energy gaugino (running) masses are determined 
with the low energy values of the SM gauge couplings 
and $f_Gm_0$: 
\dis{ \label{lowEgauginoMass}
M_a(t_T)=f_Gm_0\times g_a^2(t_T) . 
}
As discussed before, $m_0$ is determined such that 
the low energy stop masses are around $3$-$4 \tev$ 
for explaining the 126 GeV Higgs mass. 
We will discuss the valid range of $f_G$ 
in view of naturalness. 
Note that the low energy gaugino masses,  
\eq{lowEgauginoMass} are not affected by a messenger scale.

Above the messenger mass scale, however, 
the RG evolution of the MSSM gauge couplings should be 
modified by the messenger fields, 
$\{{\bf 5}_M,~\overline{\bf 5}_M\}$: 
the mGgM effects enter in the RG equations at the messenger mass scale $y_S\langle S\rangle$. 
Accordingly, all the RG evolutions of 
the MSSM Yukawa couplings and soft mass parameters 
should also be modified above the messenger scale.   

Although $y_S$ does not contribute to the soft masses 
in \eq{GGsoftmass}, it does to the messenger mass scale. 
Nonetheless, we will show later that 
the low energy mass spectra are not sensitive to $y_S$.
Since $F_S$ is proportional to $m_0$,
the MSSM gaugino masses are also proportional to $m_0$.  
As a result, they could be useful for reducing the size of 
the $m_0^2$ coefficient, and so 
for improving the fine-tuning associated with the EW scale 
and the Higgs boson mass in the mGrM or mSUGRA.  
We will discuss this issue in more detail later.



\subsection{Focus Point in the Minimal Mixed Mediation}
  
In this subsection we will discuss the focus point of $m_{h_u}^2$ 
and fine-tunings in the Minimal Mixed Mediation of SUSY breaking.      

\subsubsection{Case for $Q_M\lesssim M_{\rm GUT}$}

We first consider the case that 
the messenger mass scale is of order the GUT scale or slightly lower.  
It corresponds to the case of $|y_S|\sim {\cal O}(1)$, 
assuming $\langle S\rangle\sim {\cal O}(M_G)$. 
%
%
For simplicity, we neglect the contributions from GUT gauge multiplets such as $X$, $Y$, and their superpartners to \eq{GGsoftmass},   
since they would not much affect 
the low energy values of $\{m_{h_u}^2, m_{q_3}^2, m_{u_3^c}^2\}$  
as in the case of $|y_S\langle S\rangle|\ll {\cal O}(M_G)$. 
%
The discussion on such a relatively simple case 
is necessary also for the discussion on the case of 
$|y_S|\ll {\cal O}(1)$, i.e. the case of low messenger scale.  
As will be seen later, how small the messenger mass scale is compared to the GUT scale is indeed not very important. 
Since the gaugino masses are assumed to be generated 
dominantly by mGgM, ``$(m_{1/2}/g_0^2)$'' in 
\eqs{RG4}-(\ref{F(t)}) is just replaced by 
\dis{
\frac{m_{1/2}}{g_0^2}\approx f_Gm_0 ,
}
because they are generated around the GUT scale, $\frac{M_a(t_0)}{g_a^2(t_0)}=f_Gm_0$ ($a=3,2,1$) in this case. 
%
As a result, we can expect that in the Minimal Mixed Mediation, 
the $C_g$ terms as well as the $C_m$ terms in \eq{STmGr} 
are converted to members of $C_s$ terms.  
Since they make negative contributions to $m_{h_u}^2(t_T)$, 
they would be helpful for reducing the size of $C_s$ and 
eventually $\Delta_{m_0^2}$  \cite{old}, particularly 
in large $\tb$ cases. 

On the other hand, the soft squared masses
are induced by both the mGrM and mGgM effects at the GUT scale. 
In \eqs{Sol1}, (\ref{Sol2}), and (\ref{Sol3}), hence, 
$m_{q_30}^2$, $m_{u^c_30}^2$, $m_{h_u0}^2$, and $X_0$ 
are written down as follows: 
\begin{eqnarray}
m_{q_30}^2&\approx&m_0^2+f_G^2m_0^2
\left[\frac{8}{3}g_3^4(t_0)+\frac{3}{2}g_2^4(t_0)+\frac{1}{30}g_1^4(t_0)\right]
\approx m_0^2\left(1+\frac{21}{5}f_G^2g_0^4\right)
\\
m_{u^c_30}^2&\approx&m_0^2+f_G^2m_0^2
\left[\frac{8}{3}g_3^4(t_0)+\frac{8}{15}g_1^4(t_0)\right]
\approx m_0^2\left(1+\frac{16}{5}f_G^2g_0^4\right)
\\
m_{h_u0}^2&\approx&m_0^2+f_G^2m_0^2
\left[\frac{3}{2}g_2^4(t_0)+\frac{3}{10}g_1^4(t_0)\right]
\approx m_0^2\left(1+\frac{9}{5}f_G^2g_0^4\right)
\\
X_0&\approx&3m_0^2+f_G^2m_0^2\left[\frac{16}{3}g_3^4(t_0)+3g_2^4(t_0)+\frac{13}{15}g_1^4(t_0)\right]
\approx 3m_0^2\left(1+\frac{46}{15}f_G^2g_0^4\right)
\end{eqnarray}
For $t\leq t_0$, therefore, the semianalytic RG solutions 
\eqs{Sol1}-(\ref{Sol3}) are given as the following expressions 
in the mGgM case: 
\dis{ \label{mhuGUT}
m_{h_u}^2(t)&\approx \frac{3m_0^2}{2}\left[e^{\frac{3}{4\pi^2}\int^{t}_{t_0}dt^\prime y_t^2}-\frac13\right]+f_G^2m_0^2
\left[\frac{3}{2}g_2^4(t_0)+\frac{3}{10}g_1^4(t_0)\right]
\\
&+\frac{f_G^2m_0^2}{2}\left[\frac{16}{3}g_3^4(t_0)+3g_2^4(t_0)+\frac{13}{15}g_1^4(t_0)\right]
\left[e^{\frac{3}{4\pi^2}\int^{t}_{t_0}dt^\prime y_t^2}-1\right]
\\
&+\frac{F(t)}{2}
-f_G^2m_0^2\left[\frac32\bigg\{g_2^4(t)-g_2^4(t_0)\bigg\}
+\frac{1}{22}\bigg\{g_1^4(t)-g_1^4(t_0)\bigg\}\right]
}
and 
\bea \label{mSTopGUT}
&&\bigg\{m_{q_3}^2(t)+m_{u_3^c}^2(t)\bigg\}
\approx \frac{3m_0^2}{2}\left[e^{\frac{3}{4\pi^2}\int^{t}_{t_0}dt^\prime y_t^2}+\frac13\right]
+f_G^2m_0^2\left[
\frac{16}{3}g_3^4(t_0)+\frac{3}{2}g_2^4(t_0)+\frac{17}{30}g_1^4(t_0)\right]
 \nonumber \\
&&\qquad\qquad +\frac{f_G^2m_0^2}{2}\left[\frac{16}{3}g_3^4(t_0)+3g_2^4(t_0)+\frac{13}{15}g_1^4(t_0)\right]
\left[e^{\frac{3}{4\pi^2}\int^{t}_{t_0}dt^\prime y_t^2}-1\right]
+\frac{F(t)}{2}
\\
&&\quad\quad + f_G^2m_0^2\left[
\frac{16}{9}\bigg\{g_3^4(t)-g_3^4(t_0)\bigg\}
-\frac32\bigg\{g_2^4(t)-g_2^4(t_0)\bigg\}
-\frac{17}{198}\bigg\{g_1^4(t)-g_1^4(t_0)\bigg\}\right] ,
\nonumber 
\eea
where $F(t)$ is basically given by \eq{F(t)} except that 
$m_{1/2}/g_0^2$ should be replaced by $f_Gm_0$.  
In fact, $g_{3,2,1}^4(t_0)$ in the above equations are 
all the same as the unified gauge coupling constant $g_0^4$. 
For future convenience, however, 
we leave them as the present form. 
Note that these solutions are valid only when 
$\tb$ is small enough to neglect $y_{b,\tau}$, $A_{b,\tau}$, 
$m_{d_3^c,e_3^c,l_3,h_d}^2$, etc. 
The above semianalytic solutions admit the following 
numerical estimations: 
\dis{ \label{NumMMM1}
&\qquad~ m_{h_u}^2(t_T)\approx m_0^2\bigg[0.03 -0.52 f_G^2
-0.16 f_Ga_Y -0.11 a_Y^2\bigg] ,
\\
&\left\{m_{q_3}^2(t_T)+m_{u_3^c}^2(t_T)\right\}
\approx m_0^2\bigg[1.03+2.22 f_G^2-0.16 f_Ga_Y-0.11 a_Y^2\bigg] 
}
for $\tb=5$ and $t=t_T\approx 8.2$ ($Q_T=3.5 \tev$). 
%
%

 %
%
%
\begin{table}[!h]
\begin{center}
\begin{tabular}
{c|ccc}
\hline\hline
 {\bf Case I} & {\footnotesize $A_0=0$}  & {\footnotesize $\tb=50$}~  & {\footnotesize ${\bf \Delta_{m_0^2}=1}$}  
\\ 
 {\footnotesize ${\bf m_0^2}$} & {\footnotesize $({\bf 5.5} \tev)^2$}  & {\footnotesize $({\bf 4.5} \tev)^2$}  & {\footnotesize $({\bf 3.5} \tev)^2$}  
\\ \hline
 {\footnotesize $m_{q_3}^2(t_T)$} &  {\footnotesize $(4363 \gev)^2$}  & {\footnotesize $(3551 \gev)^2$}  & {\footnotesize $(2744 \gev)^2$} 
\\
 {\footnotesize $m_{u^c_3}^2(t_T)$} &  {\footnotesize $(3789 \gev)^2$}  & {\footnotesize $(3098 \gev)^2$}  & {\footnotesize $(2406 \gev)^2$} 
\\
 {\footnotesize ${\bf m_{h_u}^2(t_T)}$} & {\footnotesize $({\bf 431} \gev)^2$}  & {\footnotesize $({\bf 189} \gev)^2$}  & {\footnotesize $-({\bf 251} \gev)^2$}  
\\
 {\footnotesize $m_{h_d}^2(t_T)$} &  {\footnotesize $(2022 \gev)^2$}  & {\footnotesize $(1512 \gev)^2$}  & {\footnotesize $(1008 \gev)^2$}
\\ \hline\hline
 {\bf Case II} & {\footnotesize $A_0=-0.2~m_0$}  & {\footnotesize $\tb=50$}~  & {\footnotesize ${\bf \Delta_{m_0^2}=16}$}  
\\ 
 {\footnotesize ${\bf m_0^2}$} & {\footnotesize $({\bf 5.5} \tev)^2$}  & {\footnotesize $({\bf 4.5} \tev)^2$}  & {\footnotesize $({\bf 3.5} \tev)^2$}  
\\ \hline
 {\footnotesize $m_{q_3}^2(t_T)$} &  {\footnotesize $(4376 \gev)^2$}  & {\footnotesize $(3563 \gev)^2$}  & {\footnotesize $(2752 \gev)^2$} 
\\
 {\footnotesize $m_{u^c_3}^2(t_T)$} &  {\footnotesize $(3798 \gev)^2$}  & {\footnotesize $(3106 \gev)^2$}  & {\footnotesize $(2413 \gev)^2$} 
\\
 {\footnotesize ${\bf m_{h_u}^2(t_T)}$} & {\footnotesize $({\bf 539} \gev)^2$}  & {\footnotesize $({\bf 361} \gev)^2$}  & {\footnotesize $-({\bf 44} \gev)^2$}  
\\
 {\footnotesize $m_{h_d}^2(t_T)$} &  {\footnotesize $(2053 \gev)^2$}  & {\footnotesize $(1565 \gev)^2$}  & {\footnotesize $(1046 \gev)^2$}
\\ \hline\hline
 {\bf Case III} & {\footnotesize $A_0=-0.5~m_0$}  & {\footnotesize $\tb=50$}~  & {\footnotesize ${\bf \Delta_{m_0^2}=9}$}  
\\ 
 {\footnotesize ${\bf m_0^2}$} & {\footnotesize $({\bf 5.5} \tev)^2$}  & {\footnotesize $({\bf 4.5} \tev)^2$}  & {\footnotesize $({\bf 3.5} \tev)^2$}  
\\ \hline
 {\footnotesize $m_{q_3}^2(t_T)$} &  {\footnotesize $(4284 \gev)^2$}  & {\footnotesize $(3532 \gev)^2$}  & {\footnotesize $(2630 \gev)^2$} 
\\
 {\footnotesize $m_{u^c_3}^2(t_T)$} &  {\footnotesize $(3755 \gev)^2$}  & {\footnotesize $(3088 \gev)^2$}  & {\footnotesize $(2373 \gev)^2$} 
\\
 {\footnotesize ${\bf m_{h_u}^2(t_T)}$} & {\footnotesize $-({\bf 363} \gev)^2$}  & {\footnotesize $-({\bf 41} \gev)^2$}  & {\footnotesize $-({\bf 546} \gev)^2$}  
\\
 {\footnotesize $m_{h_d}^2(t_T)$} &  {\footnotesize $(1447 \gev)^2$}  & {\footnotesize $(1359 \gev)^2$}  & {\footnotesize $-(950 \gev)^2$}
\\ \hline\hline
 {\bf Case IV} & {\footnotesize $A_0=0$}  & {\footnotesize $\tb=25$}~  & {\footnotesize ${\bf \Delta_{m_0^2}=57}$}  
\\ 
 {\footnotesize ${\bf m_0^2}$} & {\footnotesize $({\bf 5.5} \tev)^2$}  & {\footnotesize $({\bf 4.5} \tev)^2$}  & {\footnotesize $({\bf 3.5} \tev)^2$}  
\\ \hline
 {\footnotesize $m_{q_3}^2(t_T)$} &  {\footnotesize $(4915 \gev)^2$}  & {\footnotesize $(4025 \gev)^2$}  & {\footnotesize $(3134 \gev)^2$} 
\\
 {\footnotesize $m_{u^c_3}^2(t_T)$} &  {\footnotesize $(3770 \gev)^2$}  & {\footnotesize $(3086 \gev)^2$}  & {\footnotesize $(2400 \gev)^2$} 
\\
 {\footnotesize ${\bf m_{h_u}^2(t_T)}$} & {\footnotesize $({\bf 152} \gev)^2$}  & {\footnotesize $-({\bf 220} \gev)^2$}  & {\footnotesize $-({\bf 293} \gev)^2$}  
\\
 {\footnotesize $m_{h_d}^2(t_T)$} &  {\footnotesize $(5057 \gev)^2$}  & {\footnotesize $(4136 \gev)^2$}  & {\footnotesize $(3215 \gev)^2$}
\end{tabular}
\end{center}\caption{Soft squared masses of the stops and Higgs bosons at $t=t_T\approx 8.2$ ($Q_T=3.5 \tev$) for various trial $m_0^2$s 
when the messenger scale is 
$Q_M\approx 1.3\times 10^{16} \gev$ with $f_G^2=0.13$ \cite{BK}. 
$\Delta_{m_0^2}$ indicates the fine-tuning measure for $m_0=4.5 \tev$ 
for each case. 
$m_{h_u}^2$s further decrease to be negative below $t=t_T$.    
The above mass spectra are generated using {\tt SOFTSUSY}.
}\label{tab:MMM}
\end{table}
%

For larger $\tb$ cases, refer to Table~\ref{tab:MMM}:  
it shows the results obtained by performing 
numerical analyses for the full RG equations 
with $\tb=50$ (Case I, II, and III) 
and $\tb=25$ (Case IV) \cite{BK}.
In all the cases, $f_G^2$ is set to be $0.13$ 
(i.e. $f_G\approx 0.36$).  
The fine-tuning measure $\Delta_{m_0^2}$ 
($\equiv \left|\frac{\partial{\rm log}m_Z^2}{\partial {\rm log}m_0^2}\right|
=\left|\frac{m_0^2}{m_Z^2}\frac{\partial m_Z^2}{\partial m_0^2}\right|$ \cite{FTmeasure}) 
listed for each case 
is indeed amazing: 
\dis{
\Delta_{m_0^2}\approx \{1,~16,~9;~57\} 
}
around $m_0^2=(4.5 \tev)^2$ 
for Case I, II, III, and IV, respectively. 
Case I in Table~\ref{tab:MMM} actually gives 
almost the minimum value of it for $\tb=50$.
$\Delta_{A_0}$ ($=\left|\frac{A_0}{m_Z^2}\frac{\partial m_Z^2}{\partial A_0}\right|$) are 
\dis{ \label{Delta_A}
\Delta_{A_0}\approx \{0, ~10, ~118; ~0\}
} 
for Case I, II, III, and IV, respectively. 
The $m_{h_u}^2$s at the stop mass scale in Table~\ref{tab:MMM} 
further decrease to be negative at the $Z$ boson mass scale 
by \eq{RGsm}. 
Using \eq{m_Z}, $|\mu|$s required for the desired value of 
$m_Z^2\approx (91 \gev)^2$ are estimated as 
\dis{
|\mu|\approx \{485 \gev, ~392 \gev, ~516 \gev; ~586 \gev\} 
}
for Case I, II, III, and IV, respectively. 
When $A_0/m_0=+0.1$, $\{\Delta_{m_0^2}, \Delta_{A_0}, |\mu|\}$ turn out to be about $\{22, 33, 569 \gev\}$. 
Therefore, we can conclude the parameter range 
\dis{
-0.5 ~<~  A_0/m_0 ~\lesssim ~ +0.1 
\quad {\rm and} \quad \tb \gtrsim 25  
}  
allows $\{\Delta_{m_0^2}, \Delta_{A_0}\}$ and $|\mu|$ 
to be smaller than $100$ 
and $600 \gev$, respectively.  
Note that $\tb=50$ is easily achieved e.g. 
from the minimal SO(10) \cite{GUT} or 
even from the MSSM embedded 
in a class of the heterotic stringy models \cite{StringModel}.   

$f_G$ is also a UV parameter in the Minimal Mixed Mediation 
and so a comment on $\Delta_{f_G}$ might be needed. 
While $\langle S\rangle$ can be fixed to be $v_G$ 
by a GUT model, $\langle F_S\rangle/m_0$ 
is associated with the vanishing C.C. 
as discussed in Section~\ref{sec:mGrM}.  
Once $\langle F_S\rangle/m_0$ is determined 
through a fine-tuning 
with other $F$-term VEVs divided by $m_0$  
and $\langle W_H\rangle/m_0$ 
such that the C.C. vanishes, 
its variation yields a nonzero C.C. 
This problem also arises even in the mGrM or mSUGRA, 
as discussed below \eq{F_a}. 
Also in the mGgM scenario, a variation of 
$\langle F_S\rangle/\langle S\rangle$ 
could give a different C.C. 
Discussions on the vanishing C.C. are  
beyond the scope of our paper.  
We will present the valid range of $f_G$ 
in Section~\ref{subsec:C}.  

With $f_G^2=0.13$ and $m_0^2=(4.5 \tev)^2$, 
\eq{lowEgauginoMass} yields 
the gluino, wino, and bino masses as follows:  
\dis{ \label{gauginoMass1}
M_{3,2,1}\approx \{1.7 \tev,~660 \gev,~360 \gev\} 
}  
for all the cases considered in Table~\ref{tab:MMM}. 
Note that they all are low energy running masses. 
The physical mass particularly for the gluino would be 
a bit heavier than it \cite{OnshellGluino}.  
Since low energy gaugino masses are not affected by a messenger scale, 
\eq{gauginoMass1} should be valid even for other choices of $y_S$.

In the above cases, the sbottom and sleptons   
turn out to be quite heavier than $3 \tev$. 
The first two generations of SUSY particles must 
be much heavier than them 
because of their extremely small relevant Yukawa couplings.     
%
%
%
Accordingly, the bino is the lightest superparticle (LSP). 
To avoid overclose of the bino dark matter in the Universe, 
some entropy production \cite{DM1} or 
other lighter dark matter such as the axino and axion is needed \cite{DM2}. 


\subsubsection{Case for $Q_M\ll M_{\rm GUT}$}

Since the mass of the messenger fields 
$\{{\bf 5}_M,\overline{\bf 5}_M\}$ is given 
by $y_S\langle S\rangle$, 
the RG evolutions of the gauge and Yukawa coupling constants 
and soft mass parameters should be modified by them 
from those of the MSSM 
above the messenger mass scale, 
$Q> y_S\langle S\rangle$. 
Although $\langle S\rangle$ can be fixed with a proper UV model, 
$y_S$ still remains as a free parameter. 
Thus, one might anticipate that low energy values of 
$m_{h_u}^2$ would be quite sensitive to $y_S$.  
In this subsection, we attempt to show that 
$\{m_{h_u}^2, m_{q_3}^2, m_{u_3^c}^2\}$ 
at the stop decoupling scale 
are very {\it insensitive} to $y_S$ 
unlike the naive expectation. 
Although we first discuss a small $\tb$ case 
for a qualitative understanding, using semianalytic expressions, 
the result is quite general: 
we will display later the numerical result for a large $\tb$ case.   
  

In the energy scale between the GUT and the messenger scales, 
only the mGrM effects are active: 
the mGgM effects come in below the messenger scale.   
Since we neglect the gaugino masses by mGrM in this paper, 
$m_{q_3}^2$, $m_{u_3^c}^2$, and $m_{h_u}^2$ for $t_M<t<t_0$ 
are simply  
\bea
&&m_{q_31}^2(t)=m_{0}^2+\frac{3m_0^2}{6}\left[e^{\frac{3}{4\pi^2}\int^{t}_{t_0}dt ~\bar{y}_t^2}-1\right] + \frac{ F_1(t)}{6} ,
\\
&&m_{u_3^c1}^2(t)=m_{0}^2+\frac{3m_0^2}{3}\left[e^{\frac{3}{4\pi^2}\int^{t}_{t_0}dt ~\bar{y}_t^2}-1\right] + \frac{F_1(t)}{3} ,
\\
&&m_{h_u1}^2(t)=m_{0}^2+\frac{3m_0^2}{2}\left[e^{\frac{3}{4\pi^2}\int^{t}_{t_0}dt ~\bar{y}_t^2}-1\right] + \frac{ F_1(t)}{2} ,
\eea
where $\bar{y}_t$ means 
the top quark Yukawa coupling constant modified
by the messenger fields for $t>t_M$. 
They can be obtained from Eqs.~(\ref{Sol1})-(\ref{Sol3}) 
and (\ref{mGrMbdy}).  
$F_1(t)$ in the above equations is obtained just by neglecting 
$m_{1/2}/g_0^2$ and setting $A_0=a_Ym_0$ in \eq{F(t)}: 
\dis{ \label{F1}
F_1(t)
=a_Y^2m_0^2 ~e^{\frac{3}{4\pi^2}\int^{t}_{t_0}dt ~\bar{y}_t^2}\left[e^{\frac{3}{4\pi^2}\int^{t}_{t_0}dt^\prime ~\bar{y}_t^2}-1\right] .
}
Hence, we have 
\dis{
X_{t1}(t)
=m_{q_31}^2(t)+m_{u_3^c1}^2(t)+m_{h_u1}^2(t)
=3m_0^2 ~e^{\frac{3}{4\pi^2}\int^{t}_{t_0}dt ~\bar{y}_t^2}+F_1(t) . 
}

At the messenger scale $t=t_M$, the mGgM effects become active: 
the additional soft masses squared, 
Eqs.~(\ref{mGgM3})-(\ref{mGgM2}), and the gaugino masses 
by \eq{GGsoftmass} should be imposed to the RG solutions, 
Eqs.~(\ref{Sol1})-(\ref{Sol3}) at $t=t_M$. 
For $t_T\leq t\leq t_M$, therefore, we get 
\begin{eqnarray}
&&m_{q_3}^2(t)=m_{q_3}^2(t_{M})+\frac{X_t(t_{M})}{6}
\left[e^{\frac{3}{4\pi^2}\int^{t}_{t_M}dt^\prime y_t^2}-1\right]
+\frac{F_2(t)}{6}
\\
&&\qquad +f_G^2m_0^2\left[\frac89\bigg\{g_3^4(t)-g_{3}^4(t_{M})\bigg\}
-\frac{3}{2}\bigg\{g_2^4(t)-g_{2}^4(t_{M})\bigg\}
-\frac{1}{198}\bigg\{g_1^4(t)-g_{1}^4(t_{M})\bigg\}\right] ,
\qquad 
\nonumber  \\
&&m_{u_3^c}^2(t)=m_{u_3^c}^2(t_{M})+\frac{X_t(t_{M})}{3}
\left[e^{\frac{3}{4\pi^2}\int^{t}_{t_M}dt^\prime y_t^2}-1\right]
+\frac{F_2(t)}{3}
\\
&&\qquad +f_G^2m_0^2\left[\frac89\bigg\{g_3^4(t)-g_{3}^4(t_{M})\bigg\}
-\frac{8}{99}\bigg\{g_1^4(t)-g_{1}^4(t_{M})\bigg\}\right] ,
\nonumber \\
&&m_{h_u}^2(t)=m_{h_u}^2(t_{M})+\frac{X_t(t_{M})}{2}
\left[e^{\frac{3}{4\pi^2}\int^{t}_{t_M}dt^\prime y_t^2}-1\right]
+\frac{F_2(t)}{2}
%
\label{mhu^2} \\
&&\qquad -f_G^2m_0^2\left[\frac32\bigg\{g_2^4(t)-g_{2}^4(t_{M})\bigg\}
+\frac{1}{22}\bigg\{g_1^4(t)-g_{1}^4(t_{M})\bigg\}\right] ,
\nonumber
\end{eqnarray}
where 
$m_{q_3}^2(t_{M})=m_{q_31}^2(t_{M})+\delta m_{u_3^c}^2|_M$, 
$m_{u_3^c}^2(t_{M})=m_{u_3^c1}^2(t_{M})+\delta m_{u_3^c}^2|_M$, 
$m_{h_u}^2(t_{M})=m_{h_u1}^2(t_{M})+\delta m_{h_u}^2|_M$, 
$X_t(t_M)=X_{t1}(t_M)+\delta X_t|_M$, etc., and so 
\begin{eqnarray}
&&m_{q_3}^2(t_{M})=\frac{3m_0^2}{2}\left[
\frac13 e^{\frac{3}{4\pi^2}\int^{t_{M}}_{t_0}dt ~\bar{y}_t^2}+\frac13\right] + \frac{ F_1(t_M)}{6} 
+ f_G^2m_0^2
\left[\frac{8}{3}g_3^4(t_M)+\frac{3}{2}g_2^4(t_M)+\frac{g_1^4(t_M)}{30}\right] ,
\qquad \\
&&m_{u_3^c}^2(t_{M})=\frac{3m_0^2}{2}\left[
\frac23 e^{\frac{3}{4\pi^2}\int^{t_{M}}_{t_0}dt ~\bar{y}_t^2}+0\right] + \frac{ F_1(t_M)}{3} 
+ f_G^2m_0^2
\left[\frac{8}{3}g_3^4(t_M)+\frac{8}{15}g_1^4(t_M)\right] ,
\\
&&m_{h_u}^2(t_{M})=\frac{3m_0^2}{2}\left[e^{\frac{3}{4\pi^2}\int^{t_{M}}_{t_0}dt ~\bar{y}_t^2}-\frac13\right] + \frac{ F_1(t_M)}{2} 
+ f_G^2m_0^2
\left[\frac{3}{2}g_2^4(t_M)+\frac{3}{10}g_1^4(t_M)\right] ,
\\
&&X_t(t_M)=3m_0^2 ~e^{\frac{3}{4\pi^2}\int^{t_{M}}_{t_0}dt ~\bar{y}_t^2}+F_1(t_M)
+f_G^2m_0^2\left[\frac{16}{3}g_3^4(t_M)+3g_2^4(t_M)+\frac{13}{15}g_1^4(t_M)\right] . 
\end{eqnarray}
Here, $g_i^4(t_M)$s ($i=3,2,1$) are extrapolated   
from their low energy values,  
using the ordinary MSSM RG equations 
without the messenger fields.  
In the above equations, $F_2(t)$ is basically given by \eq{F(t)}, but $t_0$ should be replaced by $t_M$. 
For its definition, refer to the Appendix. 

We should note that the top quark Yukawa coupling 
in the presence of the messengers 
$\{{\bf 5}_M,\overline{\bf 5}_M\}$, 
$\bar{y}(t)$ is {\it not} much different from $y_t(t)$, 
i.e. that in the absence of them above the messenger scale. 
As a result, we have  
\dis{ \label{approx1}
\frac{e^{\frac{3}{4\pi^2}\int^{t_{M}}_{t_0}dt ~\bar{y}_t^2}}
{e^{\frac{3}{4\pi^2}\int^{t_{M}}_{t_0}dt ~y_t^2}}
\approx 1.005 \quad [1.014,~1.032]
}
even for $t_M\approx 23.0$ ($Q_M\approx 1.0\times 10^{10} \gev$) 
[$t_M\approx 18.4$ ($Q_M=1.0\times 10^8 \gev$), $t_M\approx 13.8$ ($Q_M=1.0\times 10^6 \gev$)], 
namely, $y_S\sim {\cal O}(10^{-6})$ 
[${\cal O}(10^{-8})$, ${\cal O}(10^{-10})$].  
For a higher scale $t_M$, of course, 
the ratio must be closer to unity.  
With much larger $\tb$s, we get almost the same results.
From now on, thus, we will set $e^{\frac{3}{4\pi^2}\int^{t_{M}}_{t_0}dt ~\bar{y}_t^2}=e^{\frac{3}{4\pi^2}\int^{t_{M}}_{t_0}dt ~y_t^2}$, 
just when we show the insensitivity of $m_{h_u}^2(t_T)$ to $y_S$.  
Then, one can arrive at the following results: 
\dis{ \label{mhuInt}
m_{h_u}^2(t)&\approx \frac{3m_0^2}{2}\left[e^{\frac{3}{4\pi^2}\int^{t}_{t_0}dt^\prime y_t^2}-\frac13\right] + f_G^2m_0^2
\left[\frac{3}{2}g_2^4(t_M)+\frac{3}{10}g_1^4(t_M)\right]
\\
&+\frac{f_G^2m_0^2}{2}\left[\frac{16}{3}g_3^4(t_M)+3g_2^4(t_M)+\frac{13}{15}g_1^4(t_M)\right]
\left[e^{\frac{3}{4\pi^2}\int^{t}_{t_M}dt^\prime y_t^2}-1\right]
\\
&+\frac{a_Y^2m_0^2}{2}e^{\frac{3}{4\pi^2}\int^{t}_{t_0}dt^\prime y_t^2}\left[e^{\frac{3}{4\pi^2}\int^{t_{M}}_{t_0}dt ~\bar{y}_t^2}-1\right]
+\frac{F_2(t)}{2}
\\
&-f_G^2m_0^2\left[\frac32\bigg\{g_2^4(t)-g_{2}^4(t_{M})\bigg\}
+\frac{1}{22}\bigg\{g_1^4(t)-g_{1}^4(t_{M})\bigg\}\right] ,
}
and 
\bea \label{mSTopInt}
&&\bigg\{m_{q_3}^2(t)+m_{u_3^c}^2(t)\bigg\}
\approx \frac{3m_0^2}{2}\left[e^{\frac{3}{4\pi^2}\int^{t}_{t_0}dt^\prime y_t^2}+\frac13\right]+f_G^2m_0^2\left[
\frac{16}{3}g_3^4(t_M)+\frac{3}{2}g_2^4(t_M)+\frac{17}{30}g_1^4(t_M)\right]
 \nonumber \\
&&\qquad\qquad\qquad +\frac{f_G^2m_0^2}{2}\left[\frac{16}{3}g_3^4(t_M)+3g_2^4(t_M)+\frac{13}{15}g_1^4(t_M)\right]
\left[e^{\frac{3}{4\pi^2}\int^{t}_{t_M}dt^\prime y_t^2}-1\right]
\\
&&\qquad\qquad\qquad\qquad\quad ~~
+\frac{a_Y^2m_0^2}{2}e^{\frac{3}{4\pi^2}\int^{t}_{t_0}dt^\prime y_t^2}\left[e^{\frac{3}{4\pi^2}\int^{t_{M}}_{t_0}dt ~\bar{y}_t^2}-1\right]
+\frac{F_2(t)}{2}
\nonumber \\
&&\qquad\qquad + f_G^2m_0^2\left[
\frac{16}{9}\bigg\{g_3^4(t)-g_3^4(t_M)\bigg\}
-\frac32\bigg\{g_2^4(t)-g_2^4(t_M)\bigg\}
-\frac{17}{198}\bigg\{g_1^4(t)-g_1^4(t_M)\bigg\}\right] , 
\nonumber 
\eea
%
where $F_2(t)$ is recast to 
\dis{ \label{F2'}
F_2(t)\approx&\frac{f_G^2m_0^2}{64\pi^4}
\bigg[\left(e^{\frac{3}{4\pi^2}\int^{t}_{t_M}dt^\prime y_t^2}\int^{t}_{t_M}dt^\prime ~G_A ~e^{\frac{-3}{4\pi^2}\int^{t^\prime}_{t_M}dt^{\prime\prime} y_t^2}\right)^2
\\
&\qquad\qquad~~~ -2 ~e^{\frac{3}{4\pi^2}\int^{t}_{t_M}dt^\prime y_t^2}
\int^{t}_{t_M}dt^\prime ~G_A \int^{t^\prime}_{t_M}dt^{\prime\prime}
~G_A~
e^{\frac{-3}{4\pi^2}\int^{t^{\prime\prime}}_{t_M}dt^{\prime\prime\prime} y_t^2}\bigg]
\\
&-\frac{f_G^2m_0^2}{4\pi^2} \bigg[
e^{\frac{3}{4\pi^2}\int^{t}_{t_M}dt^\prime y_t^2}
\int^{t}_{t_M}dt^\prime ~G_X^2 ~e^{\frac{-3}{4\pi^2}\int^{t^\prime}_{t_M}dt^{\prime\prime} y_t^2}
-\int^{t}_{t_M}dt^\prime ~G_X^2
\bigg]
\\
&+\frac{f_Ga_Ym_0^2}{4\pi^2} ~e^{\frac{3}{4\pi^2}\int^{t}_{t_0}dt^\prime y_t^2}\left[
\int^{t}_{t_M}dt^\prime ~G_A-e^{\frac{3}{4\pi^2}\int^{t}_{t_M}dt^\prime y_t^2} \int^{t}_{t_M}dt^\prime ~G_A ~e^{\frac{-3}{4\pi^2}\int^{t^\prime}_{t_M}dt^{\prime\prime} y_t^2}
\right]
\\
&+a_Y^2m_0^2
~e^{\frac{3}{4\pi^2}\int^{t}_{t_0}dt^\prime y_t^2}
\bigg[e^{\frac{3}{4\pi^2}\int^{t}_{t_0}dt^\prime y_t^2}
-e^{\frac{3}{4\pi^2}\int^{t_{M}}_{t_0}dt^\prime \bar{y}_t^2}\bigg] . 
}
The coefficients of $a_Y^2$ in Eqs.~(\ref{mhuInt}) and (\ref{mSTopInt}) are determined from 
the third lines of them 
and the last line of \eq{F2'}:  
\dis{
\frac{a_Y^2m_0^2}{2} 
~e^{\frac{3}{4\pi^2}\int^{t}_{t_0}dt^\prime y_t^2}
\left[e^{\frac{3}{4\pi^2}\int^{t}_{t_0}dt^\prime y_t^2}-1\right] ,
}
which is coincident with those of \eqs{mhuGUT} and (\ref{mSTopGUT}). 
See the last line of \eq{F(t)}.

%
%
\begin{figure}
\begin{center}
\includegraphics[width=0.55\linewidth]{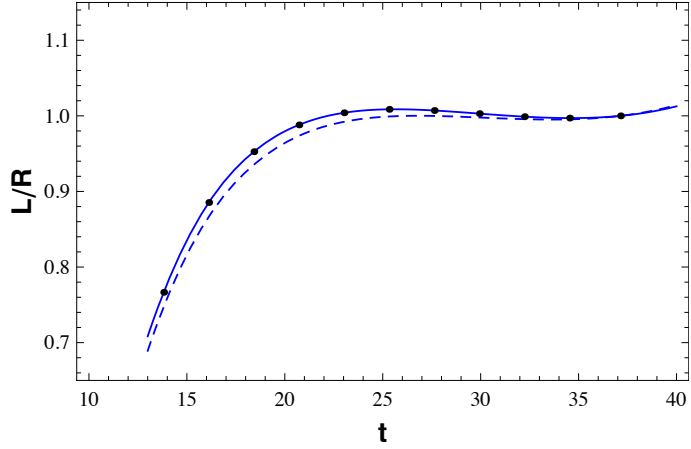}
\end{center}
\caption{Left-hand side/Right-hand side of \eq{comparison} 
vs. $t$ [$\equiv {\rm log}(Q/\gev)$]. 
The solid (dotted) line corresponds to the case of $\tb=5$ 
($\tb=50$). 
In the both cases, \eq{comparison} becomes approximately valid 
for $t\gtrsim 18.4$ [or $Q\gtrsim 10^{8} \gev$]. 
}
\label{fig:comparison}
\end{figure}
%
%

Now let us compare \eq{mhuInt} with (\ref{mhuGUT}). 
The first two terms of \eq{mhuInt} are the same as those of (\ref{mhuGUT}).  
The largest terms among the other ones would be those proportional to $g_3^4(t_M)$. 
Interestingly enough, the terms 
in the second line of the both equations are almost the same: 
\dis{ \label{comparison}
&\left[\frac{16}{3}g_3^4(t_M)+3g_2^4(t_M)+\frac{13}{15}g_1^4(t_M)\right]
\left[e^{\frac{3}{4\pi^2}\int^{t_{T}}_{t_M}dt ~y_t^2}-1\right]
\\
&\approx \left[\frac{16}{3}g_3^4(t_0)+3g_2^4(t_0)+\frac{13}{15}g_1^4(t_0)\right]
\left[e^{\frac{3}{4\pi^2}\int^{t_T}_{t_0}dt^\prime y_t^2}-1\right]
}
even for $t_M\ll t_0$. 
Fig.~\ref{fig:comparison} shows the ratio between 
the left-hand side (``L'') and the right-hand side (``R'') 
of \eq{comparison} with $t$ [$\equiv {\rm log}(Q/\gev)$]:  
\eq{comparison} becomes approximately valid 
for $t\gtrsim 18.4$ or $Q\gtrsim 10^{8} \gev$ 
regardless of the size of $\tb$.   
Note that both $g_i^4(t_M)$s in \eq{mhuInt} and 
$g_i^4(t_0)$s in \eq{mhuGUT} are determined 
from their low energy values
with the ordinary MSSM RG equations 
without the messenger fields. 

Both $g_2^4(t_M)$ and $g_1^4(t_M)$ are quite small 
for $t_M\ll t_0$.  
Since the beta function coefficient of $g_2^2(t)$ 
is still small enough ($=1$), 
$g_2^4(t_M)$ of \eq{mhuInt} is similar to 
$g_2^4(t_0)$ of \eq{mhuGUT}: 
$g_2^4(t_M)/g_2^4(t_0)$ is about $0.943$, $0.848$, and 
$0.767$ for $t_M\approx 32.2$ ($Q_M=10^{14} \gev$), 
$t_M\approx 23.0$ ($Q_M=10^{10} \gev$), and 
$t_M\approx 13.8$ ($Q_M=10^{6} \gev$), respectively.
$g_1^4(t)$ is more suppressed than $g_2^4(t)$. 
$F(t)$ and $F_2(t)$ cannot make a big difference 
between \eqs{mhuGUT} and (\ref{mhuInt}): 
although they contain $g_3^4$, $g_3^6$, etc.,  
they are suppressed with a large numbers (like $64\pi^4$) and/or 
effectively canceled each other. 
As shown before, moreover, the coefficients of $a_Y^2$ 
must be the same.  

The numerical results for the semianalytic
solutions, \eqs{mhuInt} and (\ref{mSTopInt})
are given by 
\dis{ \label{NumMMM2}
&\qquad~~ m_{h_u}^2(t_T)\approx m_0^2\bigg[0.03 -0.64 f_G^2
-0.07 f_Ga_Y-0.11 a_Y^2\bigg] ,
\\
&\left\{m_{q_3}^2(t_T)+m_{u_3^c}^2(t_T)\right\}
\approx m_0^2\bigg[1.03 + 2.73f_G^2 - 0.07f_Ga_Y - 0.11a_Y^2\bigg] 
}
for $\tb=5$ and $t_M\approx 23.0$ ($Q_M=10^{10} \gev$).
%
%
The main difference in $m_{h_u}^2(t_T)$s 
of \eqs{mhuGUT} and (\ref{mhuInt}) arises from 
the difference between $g_2^4(t_0)$ and $g_2^4(t_M)$: 
\dis{
\Delta m_{h_u}^2(t_T)\approx  f_G^2 m_0^2\times 3\left[
g_2^4(t_0)-g_2^4(t_M)\right]\approx  f_G^2 m_0^2\times 0.10 ,
}
which is approximately the difference 
between \eqs{NumMMM1} and (\ref{NumMMM2}).
Similarly, the main difference 
in $\{m_{q_3}^2(t_T)+m_{u_3^c}^2(t_T)\}$ 
comes from the $f_G^2m_0^2$ parts in the first and last lines 
of \eqs{mSTopGUT} and (\ref{mSTopInt}). 
Considering the extremely large energy scale difference 
between the GUT and $10^{10} \gev$, 
the differences in \eqs{NumMMM1} and (\ref{NumMMM2}) 
are quite small. 
Moreover, such differences become more negligible 
for a small enough $f_G^2$ [$\sim {\cal O}(0.1)$]. 
Actually, we need such a small $f_G^2$ 
also to suppress the $m_0^2$ dependence of $m_{h_u}^2(t_T)$.

%
%
\begin{figure}
\begin{center}
\includegraphics[width=0.6\linewidth]{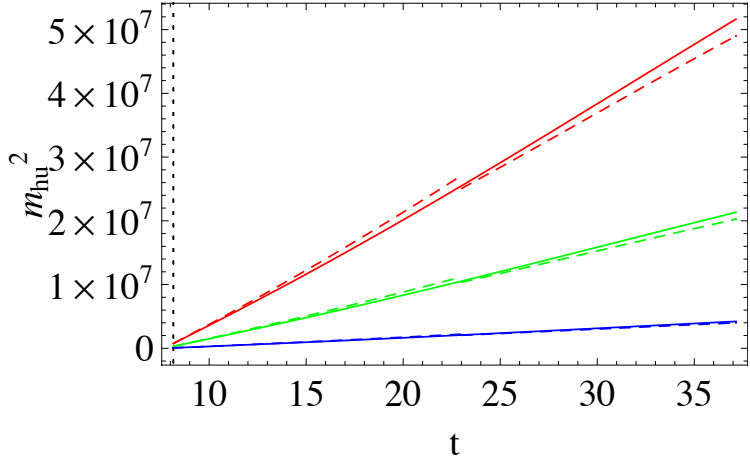}
\end{center}
\caption{RG evolutions of $m_{h_u}^2$ 
with $t$ [$\equiv {\rm log}(Q/\gev)]$
for $m_0^2=(7 \tev)^2$ [Red], $(4.5 \tev)^2$ [Green], 
and $(2 \tev)^2$ [Blue] when $f_G^2=0.13$, $A_0=-0.2 ~m_0$, 
and $\tb=50$ \cite{BK}. 
The tilted solid [dotted] lines correspond to the case of 
$t_M\approx 37$ (or $Q_M\approx 1.3\times 10^{16} \gev$, ``{\rm Case A}'') 
[$t_M\approx 23$ (or $Q_M=1.0\times 10^{10} \gev$, ``{\rm Case B}'')].
The vertical dotted line at $t=t_T\approx 8.2$ ($Q_T=3.5 \tev$) indicates 
the desired stop decoupling scale. 
The discontinuities of $m_{h_u}^2(t)$ should appear at the messenger scales.
%
%
}
\label{fig:MMM}
\end{figure}
%
%

Fig.~\ref{fig:MMM} exhibits some RG evolutions of $m_{h_u}^2$ 
under various trial $m_0^2$ when $f_G^2=0.13$, 
$A_0=-0.2m_0$, and $\tb=50$ \cite{BK}. 
The solid lines [dotted] lines correspond to the case of 
$t_M\approx 37$ (or $Q_M\approx 1.3\times 10^{16} \gev$, ``{\rm Case A}'') 
[$t_M\approx 23$ (or $Q_M=1.0\times 10^{10} \gev$, ``{\rm Case B}'')].
Since the soft masses induced by the mGgM effect are added 
at the messenger scale, 
the discontinuities of $m_{h_u}^2(t)$ should arise there.  
As seen in Fig.~\ref{fig:MMM}, 
in the case of the Minimal Mixed Mediation, 
the FP of $m_{h_u}^2$ always appears 
at the desired stop mass scale ($t=t_T\approx 8.2$)
regardless of the messenger scales:
the FP scale is not affected by messenger scales 
or the size of $y_S$. 
As defined in Section~\ref{sec:mGrM}, in fact, $m_0$ is 
originally a parameter associated with the VEV of 
the Hidden sector superpotential, $\langle W_H\rangle$, 
which triggers SUSY breaking in the observable sector,  
via both the gravity and gauge mediations, 
determining the soft mass spectrum. 
Hence, the low energy value of $m_{h_u}^2$ can remain  
insensitive to the scale of $\langle W_H\rangle$ 
and the coupling strength to the hidden sector:  
the wide ranges of UV parameters can allow almost the same $m_{h_u}^2$s at low energy.  
Under this situation, one can guess that $m_0^2\approx (4.5 \tev)^2$ happens to be selected by Nature, 
yielding $3$-$4 \tev$ stop mass and 
eventually also the $126 \gev$ Higgs mass.  
As mentioned above, 
the gaugino masses are also not affected by a messenger scale. 
In the both cases of Fig.~\ref{fig:MMM}, thus, 
the gaugino masses are given by \eq{gauginoMass1}.

\subsection{Gluino Mass Bound} \label{subsec:C}

\begin{figure}[tbp]
\centering
\subfloat[]{%
\includegraphics[width=0.5\textwidth]{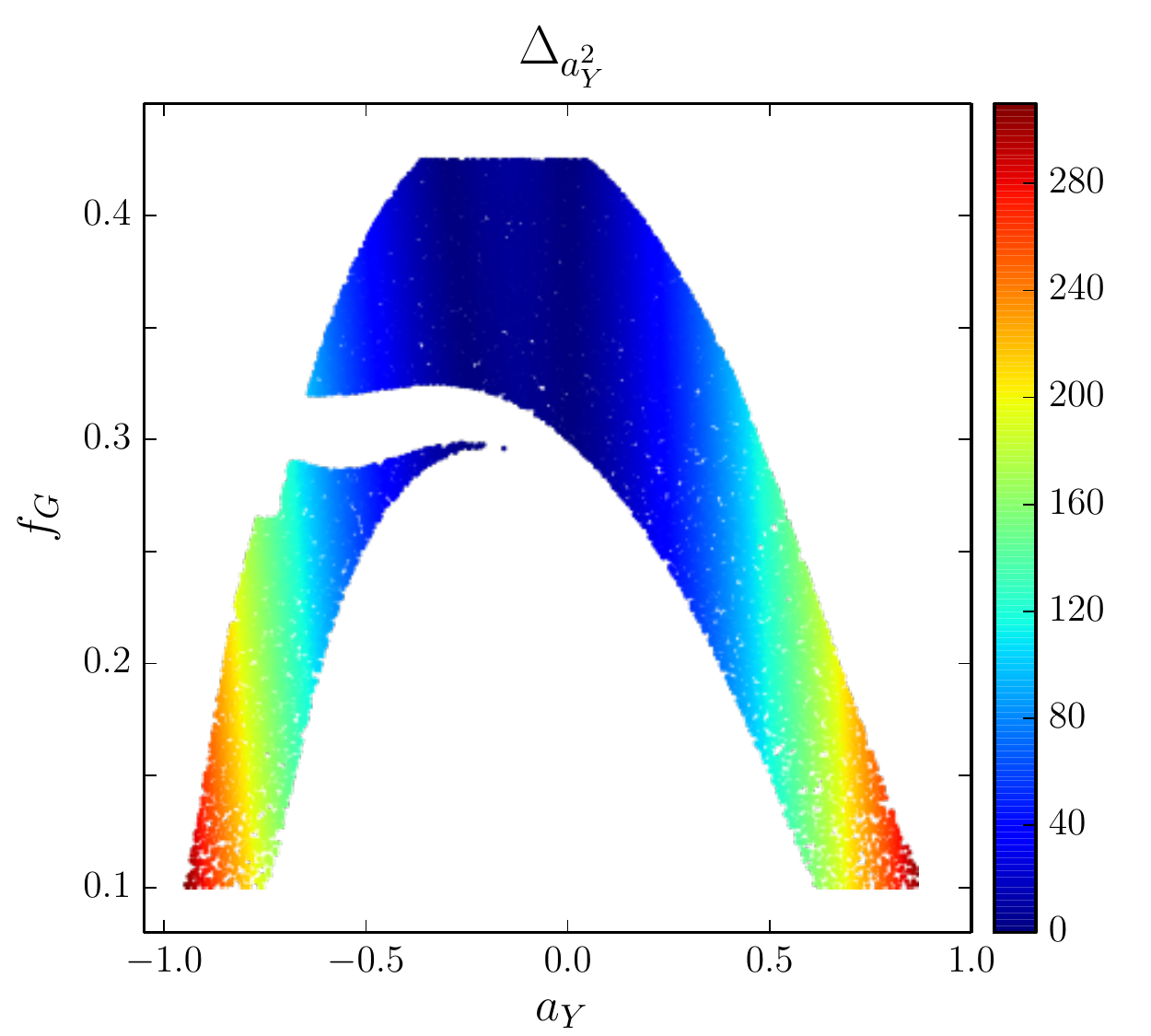}
\label{aY4000}
}%
\subfloat[]{%
\includegraphics[width=0.5\textwidth]{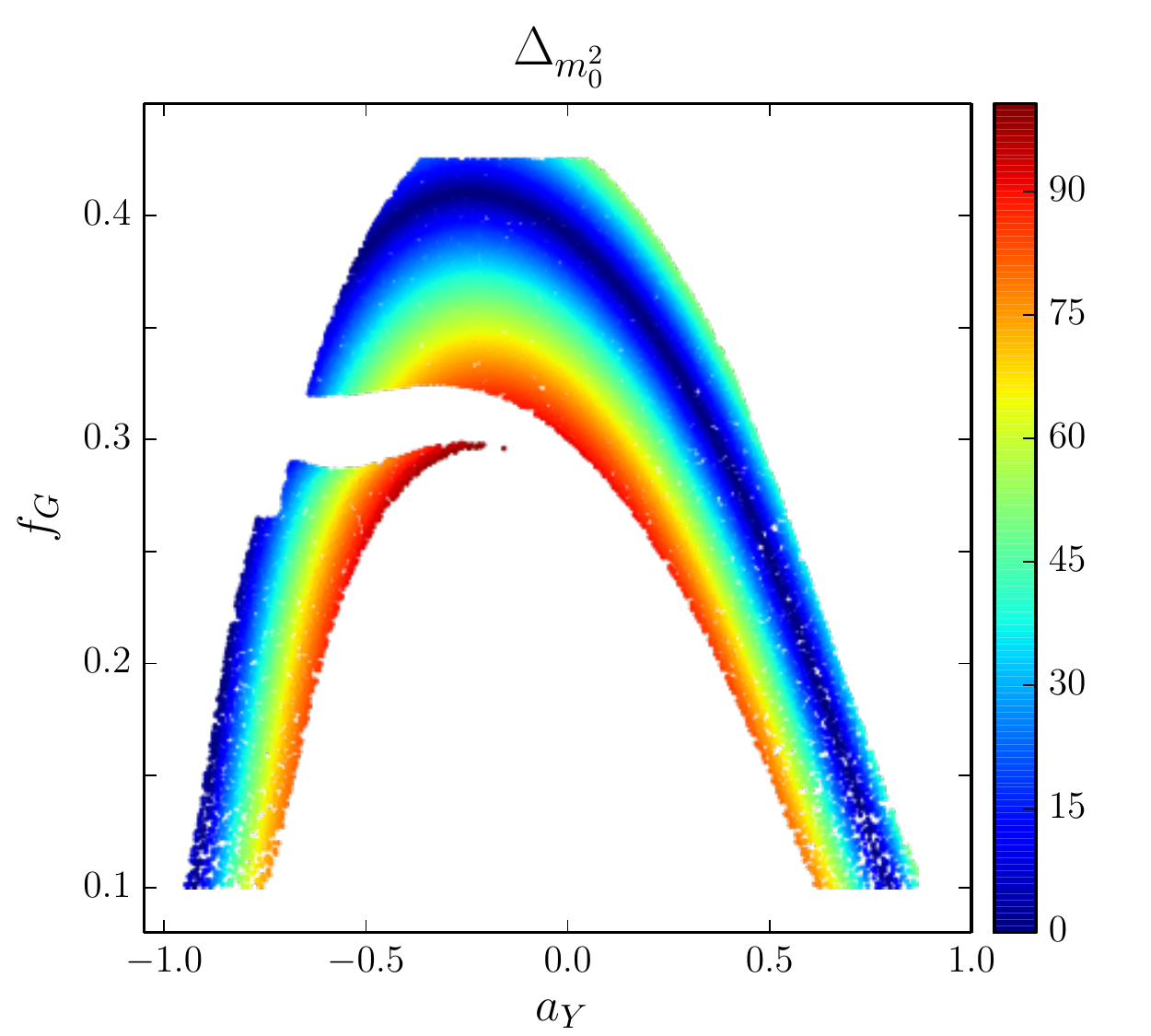}%
}%
\\
\subfloat[]{%
\includegraphics[width=0.5\textwidth]{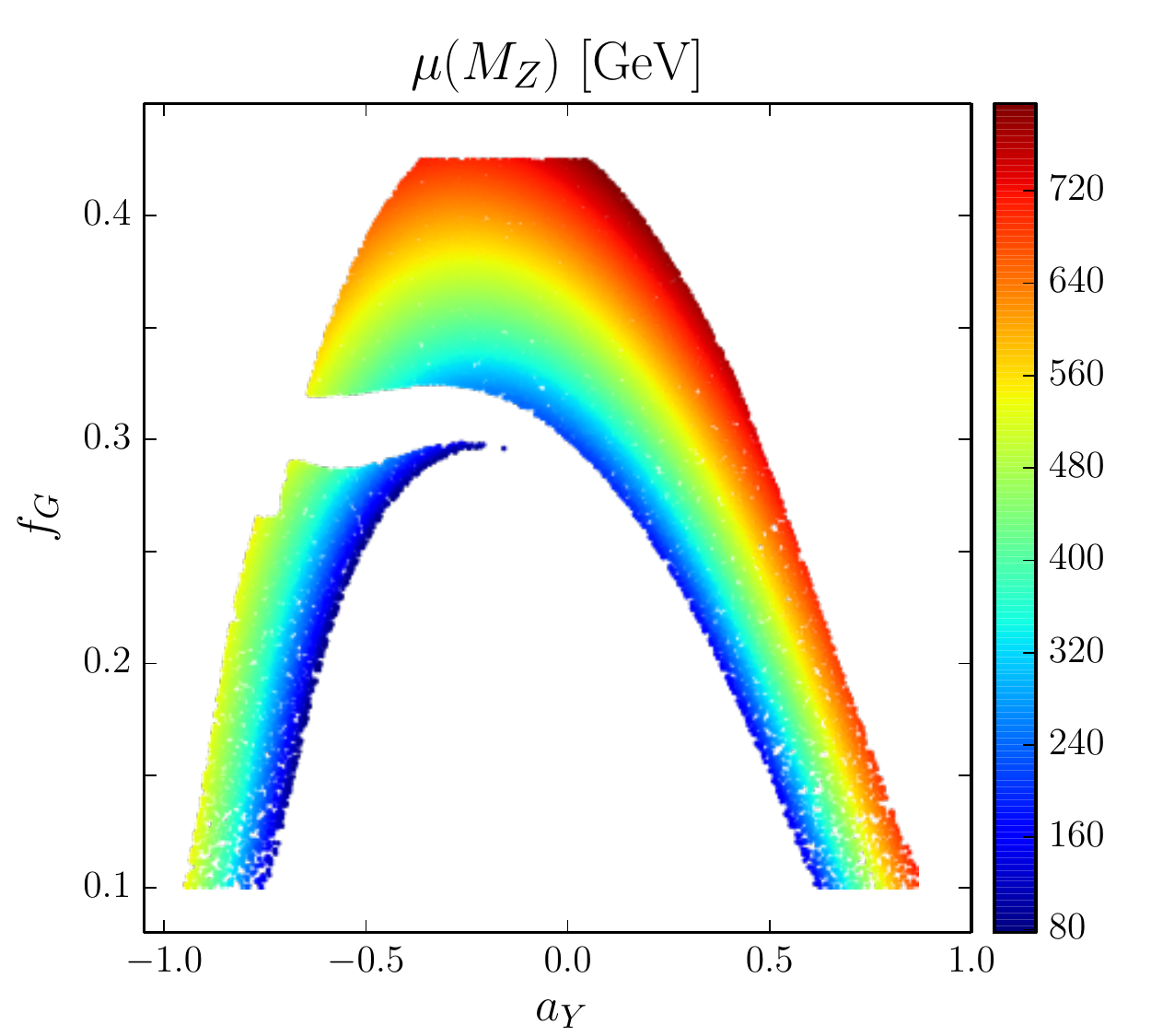}%
}%
\subfloat[]{%
\includegraphics[width=0.5\textwidth]{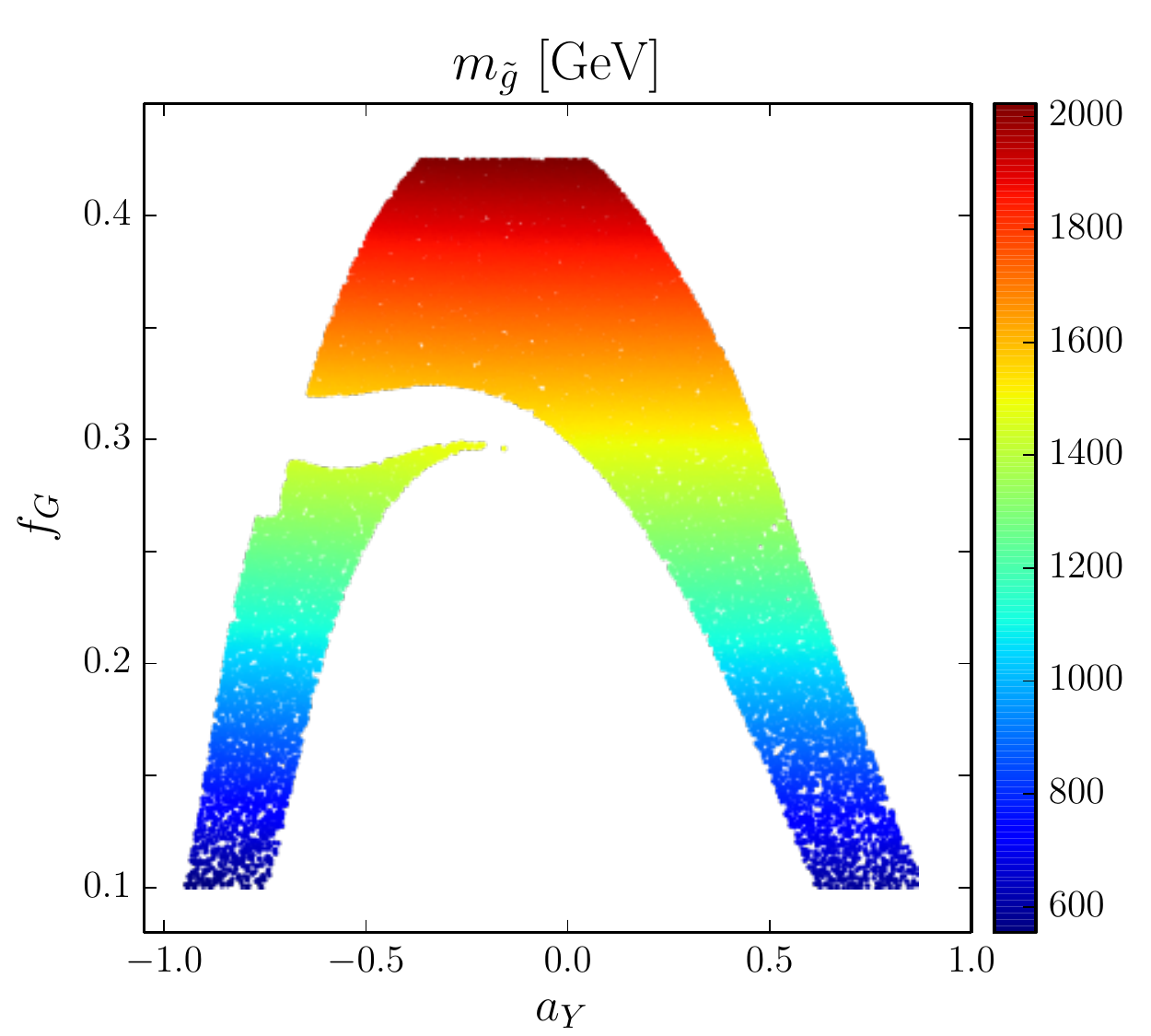}
}%
\caption[]{Scatter plots for {\bf (a)} $\Delta_{a_Y}$, {\bf (b)} $\Delta_{m_0^2}$, and
{\bf (c)} $|\mu|$ at the $M_Z$ scale,
and {\bf (d)} physical gluino mass
when $m_0^2=(4~\tev)^2$ and $\tb=50$.
The stop mass scale is about $3.0 \tev$.}
\label{fig:gluinoM4000}
\end{figure}
\begin{figure}[tbp]
\centering
\subfloat[]{%
\includegraphics[width=0.5\textwidth]{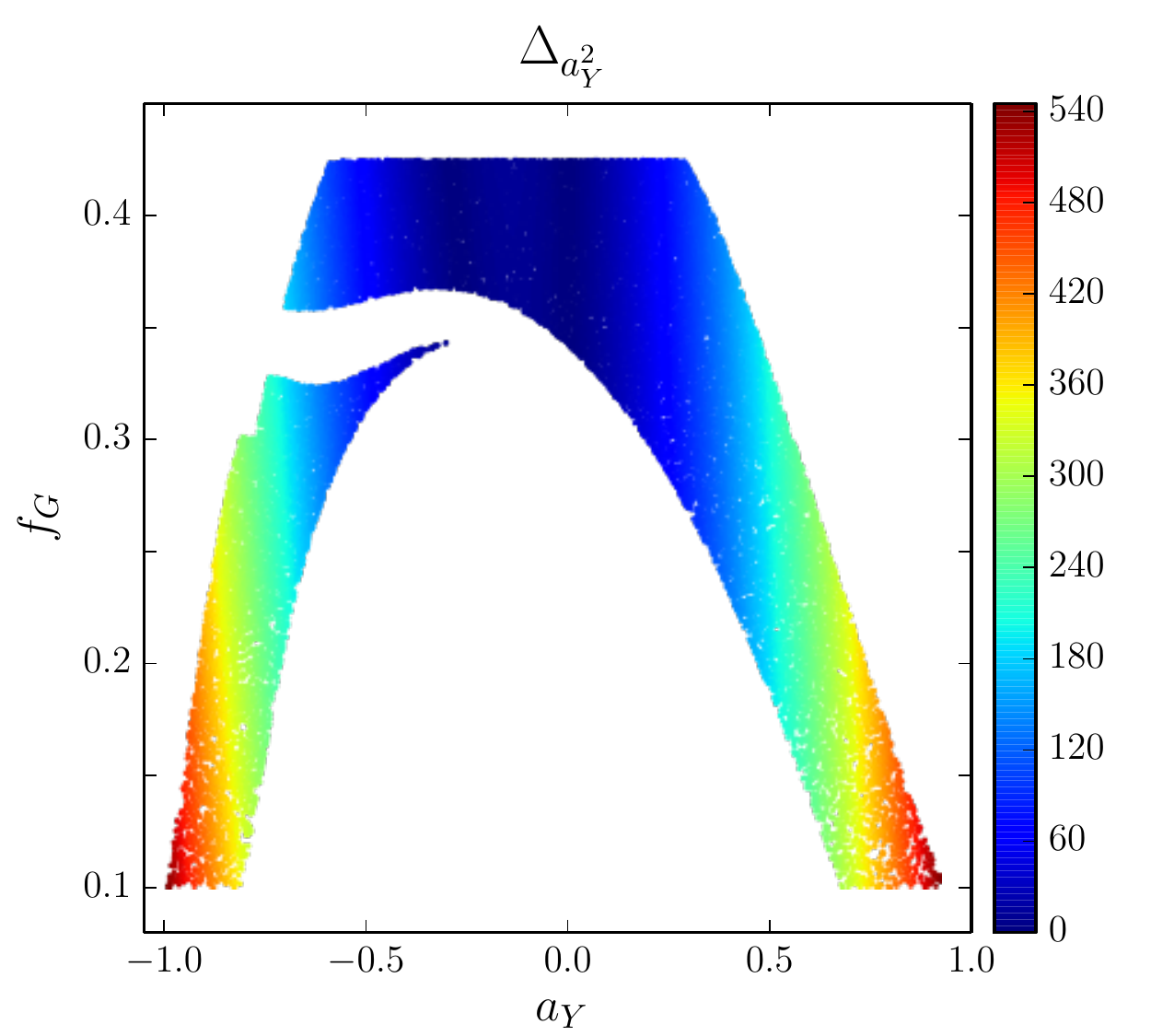}
\label{aY4000}
}%
\subfloat[]{%
\includegraphics[width=0.5\textwidth]{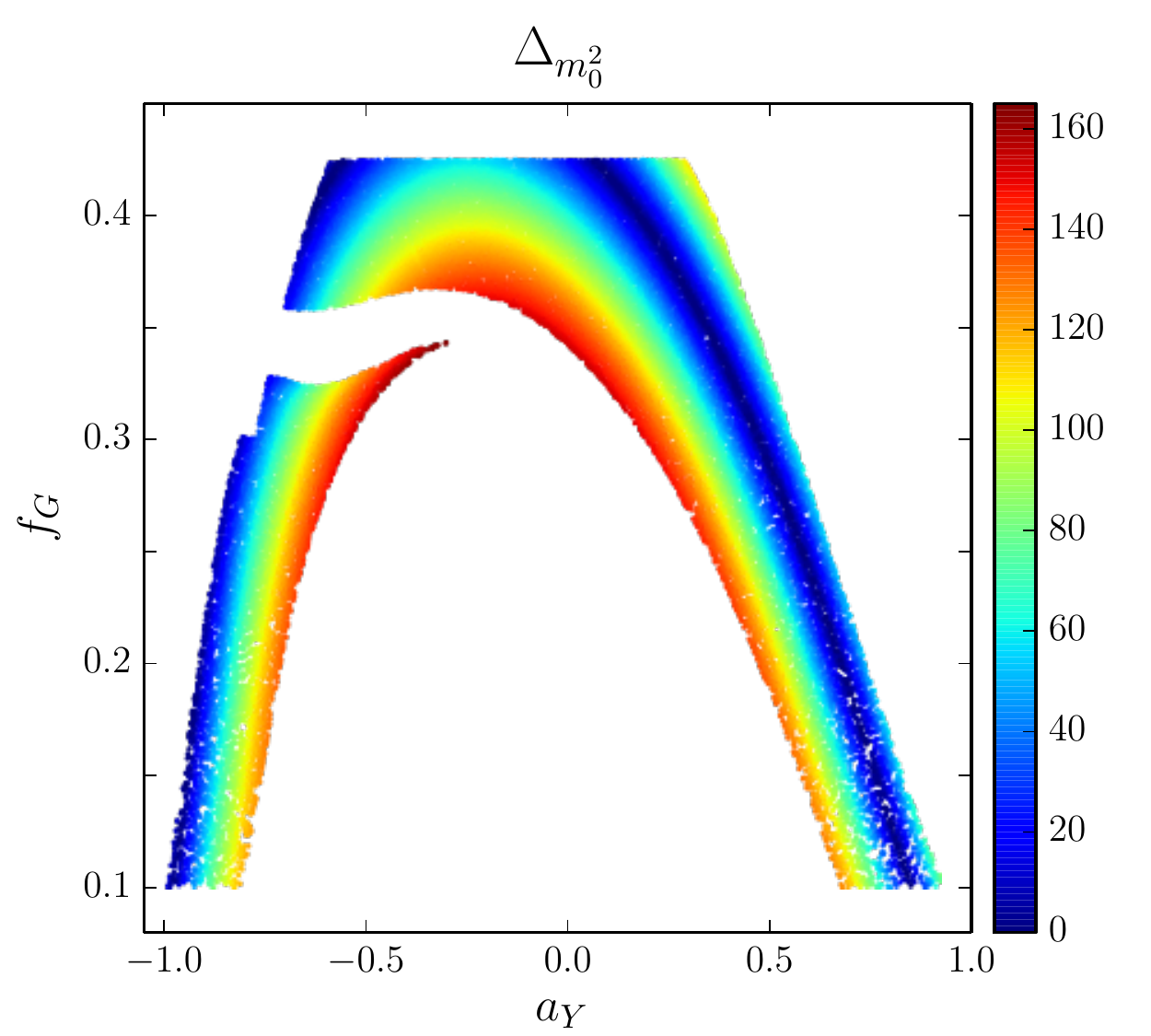}%
}%
\\
\subfloat[]{%
\includegraphics[width=0.5\textwidth]{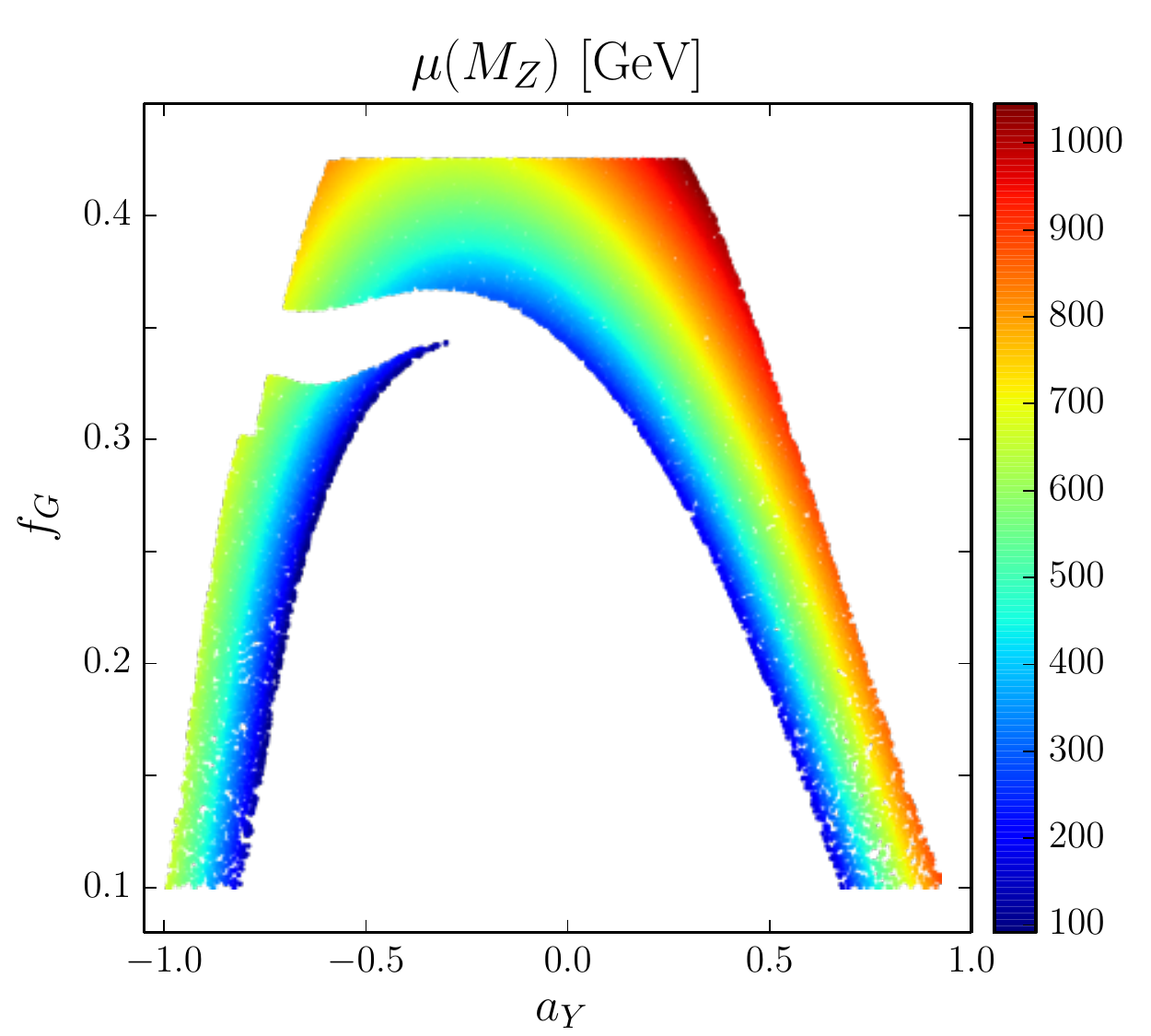}%
}%
\subfloat[]{%
\includegraphics[width=0.5\textwidth]{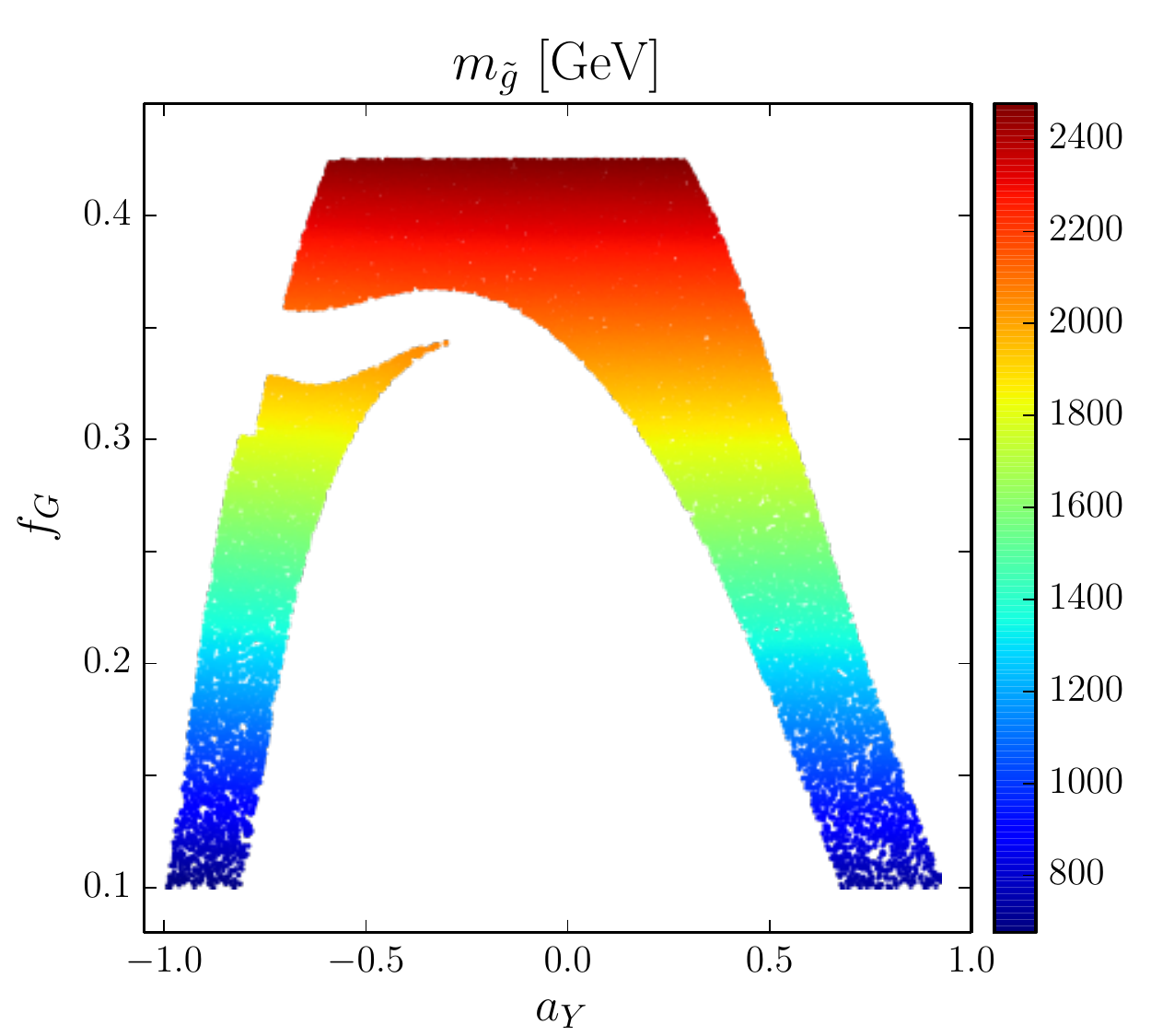}
}%
\caption[]{Scatter plots for {\bf (a)} $\Delta_{a_Y}$, {\bf (b)} $\Delta_{m_0^2}$, and
{\bf (c)} $|\mu|$ at the $M_Z$ scale,
and {\bf (d)} physical gluino mass
when $m_0^2=(5~\tev)^2$ and $\tb=50$.
The stop mass scale is about $3.7 \tev$.}
\label{fig:gluinoM5000}
\end{figure}

Fig.s~\ref{fig:gluinoM4000} and \ref{fig:gluinoM5000} show various scatter plots for given ranges of $\{f_G,~a_Y\}$ 
with $\tb=50$. 
$m_0^2$ in Fig.s~\ref{fig:gluinoM4000} and \ref{fig:gluinoM5000} are taken, respectively, to be $(4 \tev)^2$ and $(5 \tev)^2$. 
As a result, the stop mass scales are about $3.0$ and $3.7 \tev$, respectively.
Here we set $M_G$ as the scale where the EW gauge couplings, 
$g_2$ and $g_1$ meet. 
It is approximately $1.7\times 10^{16}~\gev$ in these cases.
They all are drawn using {\tt SOFTSUSY}-3.6.2. 
%
As expected from \eqs{NumMMM1} and (\ref{NumMMM2}), 
they have ``rainbow'' shapes.  
The two ``legs'' of the ``rainbow'' in those figures, 
which are located in the left and right sides for the figures, 
are relatively narrow. 
Note that the origin of disconnected points on the left legs is the convergence problem of the iterations of the {\tt SOFTSUSY} calculation. Their colors are, therefore, supposed to be interpolated continuously since they are not physically forbidden.

As $a_Y$ (or $A_0/m_0$) is deviated from zero 
$m_{h_u}^2$ is expected to rapidly change from \eqs{NumMMM1} and (\ref{NumMMM2}). 
Accordingly, $m_Z^2$ would also rapidly change.   
It implies that $\Delta_{a_Y}$ would rapidly increase 
as shown in Fig.s~\ref{fig:gluinoM4000} and \ref{fig:gluinoM5000} - {\bf (a)}, which was seen also in \eq{Delta_A}. 
For a small enough $\Delta_{a_Y}$, thus, 
we are more interested in the thick central parts 
around $a_Y=0$ in the figures,
\dis{
-0.7 ~\lesssim~ a_Y ~\lesssim~ 0.5 , 
 } 
which satisfies $\Delta_{a_Y}<100$.  
As discussed before, in addition, 
we confine our discussion to cases of $|\mu|<600 \gev$. 
In fact, the constraint associated with $\mu$ or heavy gluino effects 
could be relaxed by assuming very heavy masses 
for the superpartners of the first and second generations 
of the SM chiral fermions \cite{KS}. 
For simplicity, however, we don't consider such a possibility 
in this paper. 
Below $f_G\approx 0.3$, the EW symmetry breaking does not occur.  
From Fig.s~\ref{fig:gluinoM4000} and \ref{fig:gluinoM5000} - {\bf (c)}, thus, $f_G$ is constrained to 
\dis{
0.3 ~\lesssim~ f_G ~\lesssim~ 0.4 , 
}
which is consistent with $\Delta_{m_0^2}<100$ as seen in Fig.s~\ref{fig:gluinoM4000} and \ref{fig:gluinoM5000} - {\bf (b)}.   
From Fig.s~\ref{fig:gluinoM4000} and \ref{fig:gluinoM5000} - {\bf (d)}, we see that 
the above ranges confine the physical gluino mass to 
\dis{
1.6 \tev ~\lesssim ~&m_{\tilde{g}}~ \lesssim~ 2.2 \tev .
}        
Note that this gluino mass bound is a theoretical constraint
obtained by considering the naturalness of the EW scale 
in the  Minimal Mixed Mediation scenario. 
It is well inside the discovery potential range of LHC Run~II. 
Actually the relevant energy scale for 
the naturalness of the low energy SUSY 
in the Minimal Mixed Mediation scenario
was outside the range of LHC Run~I, 
but it can be covered by LHC Run~II.    
Accordingly, the future exploration for 
the SUSY particle, particularly, the gluino 
at the LHC would be more important.  



\section{Conclusion} \label{sec:conclusion}

In this paper, we have studied the SUSY breaking effects by  
the mGrM parametrized with $m_0$, combined with  
the mGgM parametrized with $f_G\cdot m_0$ 
for a common SUSY breaking source at a hidden sector, 
$\langle W_H\rangle$ ($\sim m_0M_P^2$) in a SUGRA framework.   
When the minimal K${\rm \ddot{a}}$hler potential and 
the minimal gauge kinetic function ($=\delta_{ab}$) 
are employed at tree level, 
a FP of $m_{h_u}^2$ appears a bit higher energy scale 
than $m_Z$ (``shifted FP''), depending on $f_G$. 
Basically $f_G$ is a parameter determined by a model.   
For $0.3\lesssim f_G \lesssim 0.4$, the FP of $m_{h_u}^2$ 
emerges at $3$-$4 \tev$ scale, which is the stop mass scale 
desired for explaining the $125 \gev$ Higgs mass, 
and so $m_{h_u}^2$ becomes quite insensitive to stop masses 
or $m_0^2$. 
Thus, this range of $f_G$ and $-0.7 \lesssim a_Y \lesssim 0.3 $ 
can admit the fine-tuning measures and $\mu$ to be 
much smaller than $100$ and $600 \gev$, respectively. 
The range $0.3\lesssim f_G \lesssim 0.4$ 
is directly translated into e.g. the gluino mass bound, 
$1.6 \tev \lesssim m_{\tilde{g}} \lesssim 2.2 \tev$, 
which could readily be tested at LHC Run~II in the near future. 


\acknowledgments

D.K. is supported by the Basic Science Research Program through the National Research Foundation of Korea (NRF) funded by the Ministry of Science, ICT \& Future Planning, Grant No. 2015R1C1A1A02037830, and also by Gyeongsangbuk-Do and Pohang City for Independent Junior Research Groups at the Asia Pacific Center for Theoretical Physics. 
B.K. is supported by 
the National Research Foundation of Korea (NRF) funded by the Ministry of Education, Grant No. 2013R1A1A2006904, 
and also in part 
by Korea Institute for Advanced Study (KIAS) grant funded by the Korean government.


\section{Appendix} \label{sec:appendix}




We present our semianalytic solutions to the RG equations. 
When ${\rm tan}\beta$ is small enough and the RH neutrinos are decoupled, 
the RG evolutions of the soft mass parameters, 
$m_{q_3}^2$, $m_{u^c_3}^2$, $m_{h_u}^2$, and $A_t$ are simplified approximately as 
\begin{eqnarray}
16\pi^2\frac{dm_{q_3}^2}{dt}&=&2y_t^2\left(X_t+A_t^2\right)-\frac{32}{3}g_3^2M_3^2-6g_2^2M_2^2-\frac{2}{15} g_1^2M_1^2 ,
\label{apdxRG1} \\
16\pi^2\frac{dm_{u^c_3}^2}{dt}&=&4y_t^2\left(X_t+A_t^2\right)-\frac{32}{3}g_3^2M_3^2-\frac{32}{15} g_1^2M_1^2 ,
\label{apdxRG2} \\
16\pi^2\frac{dm_{h_u}^2}{dt}&=&6y_t^2\left(X_t+A_t^2\right)-6g_2^2M_2^2-\frac65 g_1^2M_1^2 ,
\qquad~~ \label{apdxRG3} \\
8\pi^2\frac{dA_t}{dt}&=&6y_t^2A_t-\frac{16}{3}g_3^2M_3-3g_2^2M_2-\frac{13}{15} g_1^2M_1 
\equiv 6y_t^2A_t 
-\left(\frac{m_{1/2}}{g_0^2}\right)G_A ,
\label{apdxRG4}
\end{eqnarray} 
assuming 
$\frac{M_a(t)}{g_a^2(t)}=\frac{m_{1/2}}{g_0^2}$ ($a=3,2,1$). 
Summation of Eqs.~(\ref{apdxRG1}), (\ref{apdxRG2}), and (\ref{apdxRG3}) yields the RG equation for $X_t$ 
($\equiv m_{q_3}^2+m_{u^c_3}^2+m_{h_u}^2$):
\dis{ \label{apdxX}
\frac{dX_t}{dt} = \frac{3y_t^2}{4\pi^2}\left(X_t + A_t^2\right)
-\frac{1}{4\pi^2}\left(\frac{m_{1/2}}{g_0^2}\right)^2
G_X^2 .
}
In Eqs.~(\ref{apdxRG4}) and (\ref{apdxX}), $G_A$ and $G_X^2$ are defined in \eq{G_A}.   
%
The solutions of $A_t$ and $X_t$ are given by 
\begin{eqnarray}
&&\qquad~~ A_t(t)=e^{\frac{3}{4\pi^2}\int^t_{t_0}dt^\prime y_t^2} 
\left[A_0
-\frac{1}{8\pi^2}\left(\frac{m_{1/2}}{g_0^2}\right)
\int^t_{t_0}dt^\prime G_A
e^{\frac{-3}{4\pi^2}\int^{t'}_{t_0}dt^{\prime\prime} y_t^2}
\right] , 
\label{apdxSolA}
\\
&&X_t(t)=e^{\frac{3}{4\pi^2}\int^t_{t_0}dt^\prime y_t^2}
\left[X_0+\int^t_{t_0}dt^\prime 
\left\{\frac{3}{4\pi^2}y_t^2A_t^2
-\frac{1}{4\pi^2}\left(\frac{m_{1/2}}{g_0^2}\right)^2
G_X^2\right\}
e^{\frac{-3}{4\pi^2}\int^{t'}_{t_0}dt^{\prime\prime}y_t^2}
\right] ,
\label{apdxSolX}
\end{eqnarray}
where $A_0$ and $X_0$ denote the GUT scale values of $A_t$ and $X_t$, $A_0\equiv A_t(t=t_0)$, and $X_0\equiv X_t(t=t_0)=m_{q_30}^2+m_{u^c_30}^2+m_{h_u0}^2$.

With Eqs.~(\ref{apdxX}) and (\ref{apdxSolX}), one can solve Eqs. (\ref{apdxRG1}), (\ref{apdxRG2}), and (\ref{apdxRG3}):
\begin{eqnarray}
&&m_{q_3}^2(t)=m_{q_30}^2+\frac{X_0}{6}\left[e^{\frac{3}{4\pi^2}\int^{t}_{t_0}dt^\prime y_t^2}-1\right]
+\frac{F(t)}{6}
\label{apdxSol1} \\ \nonumber 
&&\qquad +\left(\frac{m_{1/2}}{g_0^2}\right)^2
\left[\frac89\bigg\{g_3^4(t)-g_0^4\bigg\}
-\frac32\bigg\{g_2^4(t)-g_0^4\bigg\}-\frac{1}{198}\bigg\{g_1^4(t)-g_0^4\bigg\}\right] ,
\\
%
&&m_{u^c_3}^2(t)=m_{u^c_30}^2+\frac{X_0}{3}\left[e^{\frac{3}{4\pi^2}\int^{t}_{t_0}dt^\prime y_t^2}-1\right]
+\frac{F(t)}{3}
\label{apdxSol2} \\ \nonumber 
&&\qquad +\left(\frac{m_{1/2}}{g_0^2}\right)^2
\left[\frac89\bigg\{g_3^4(t)-g_0^4\bigg\}
-\frac{8}{99}\bigg\{g_1^4(t)-g_0^4\bigg\}\right] , 
\\
&&m_{h_u}^2(t)=m_{h_u0}^2+\frac{X_0}{2}\left[e^{\frac{3}{4\pi^2}\int^{t}_{t_0}dt^\prime y_t^2}-1\right]
+\frac{F(t)}{2}
\label{apdxSol3} \\ \nonumber 
&&\qquad -\left(\frac{m_{1/2}}{g_0^2}\right)^2 
\left[\frac32\bigg\{g_2^4(t)-g_0^4\bigg\}
+\frac{1}{22}\bigg\{g_1^4(t)-g_0^4\bigg\}\right] ,
\end{eqnarray}
where $F(t)$ is defined as 
\dis{ \label{apdxF}
&\qquad\qquad F(t)\equiv 
e^{\frac{3}{4\pi^2}\int^{t}_{t_0}dt^\prime y_t^2} 
\int^{t}_{t_0}dt^\prime ~\frac{3}{4\pi^2}y_t^2A_t^2 ~e^{\frac{-3}{4\pi^2}\int^{t^\prime}_{t_0}dt^{\prime\prime}y_t^2}
\\
& -\frac{1}{4\pi^2}\left(\frac{m_{1/2}}{g_0^2}\right)^2
\left[e^{\frac{3}{4\pi^2}\int^{t}_{t_0}dt^\prime y_t^2} \int^{t}_{t_0}dt^\prime ~G_X^2 ~e^{\frac{-3}{4\pi^2}\int^{t'}_{t_0}dt^{\prime\prime}y_t^2}
-\int^{t}_{t_0}dt^\prime~G_X^2 \right] .
}
%
Using \eq{gaugeSol}, one can obtain the following results: 
\begin{eqnarray}
&& \int^t_{t_0}dt^\prime g_i^2M_i^2=\frac{4\pi^2}{b_i}\left(\frac{m_{1/2}}{g_0^2}\right)^2\left\{g_i^4(t)-g_0^4\right\} ,
\\
&& \int^t_{t_0}dt^\prime g_i^2M_i=\frac{8\pi^2}{b_i}\left(\frac{m_{1/2}}{g_0^2}\right)\left\{g_i^2(t)-g_0^2\right\} ,
\\
&& \int^t_{t_0}dt^\prime g_i^4=\frac{8\pi^2}{b_i}\left\{g_i^2(t)-g_0^2\right\} .   
\end{eqnarray} 
which are useful to get the solutions, Eqs.~(\ref{apdxSol1}), (\ref{apdxSol2}), and  (\ref{apdxSol3}).

With \eq{apdxSolA} the first line of \eq{apdxF} is recast to 
\dis{ \label{apdxF2}
&A_t^2(t)-
e^{\frac{3}{4\pi^2}\int^{t}_{t_0}dt^\prime y_t^2}\bigg\{
A_0^2 -\frac{1}{4\pi^2}\left(\frac{m_{1/2}}{g_0^2}\right)
\int^{t}_{t_0}dt^\prime ~G_A(t^\prime)A_t(t^\prime) ~e^{\frac{-3}{4\pi^2}\int^{t^\prime}_{t_0}dt^{\prime\prime} y_t^2}\bigg\}
\\
&\quad =\frac{1}{64\pi^4}\left(\frac{m_{1/2}}{g_0^2}\right)^2\bigg[
\left(e^{\frac{3}{4\pi^2}\int^{t}_{t_0}dt^\prime y_t^2}\int^{t}_{t_0}dt^\prime ~G_A ~e^{\frac{-3}{4\pi^2}\int^{t^\prime}_{t_0}dt^{\prime\prime} y_t^2}\right)^2
\\
&\qquad \qquad\qquad\qquad\quad -2 ~e^{\frac{3}{4\pi^2}\int^{t}_{t_0}dt^\prime y_t^2}
\int^{t}_{t_0}dt^\prime ~G_A \int^{t^\prime}_{t_0}dt^{\prime\prime}
~G_A~
e^{\frac{-3}{4\pi^2}\int^{t^{\prime\prime}}_{t_0}dt^{\prime\prime\prime} y_t^2}\bigg]
\\
&\quad +\frac{A_0}{4\pi^2}\left(\frac{m_{1/2}}{g_0^2}\right) ~e^{\frac{3}{4\pi^2}\int^{t}_{t_0}dt^\prime y_t^2}\left[
\int^{t}_{t_0}dt^\prime ~G_A-e^{\frac{3}{4\pi^2}\int^{t}_{t_0}dt^\prime y_t^2} \int^{t}_{t_0}dt^\prime ~G_A ~e^{\frac{-3}{4\pi^2}\int^{t^\prime}_{t_0}dt^{\prime\prime} y_t^2}
\right]
\\
&\quad +A_0^2
~e^{\frac{3}{4\pi^2}\int^{t}_{t_0}dt^\prime y_t^2}
\bigg[e^{\frac{3}{4\pi^2}\int^{t}_{t_0}dt^\prime y_t^2}-1\bigg] . 
%
}


When the gaugino masses are generated below $t_M$ with 
$\frac{M_a(t_M)}{g_a^2(t_M)}=f_Gm_0$ ($a=3,2,1$),
the solutions for $t_f<t_i<t_M$ are 
\begin{eqnarray}
&&m_{q_3}^2(t_{f})=m_{q_3}^2(t_{i})+\frac16\bigg\{X_t(t_{f})-X_t(t_{i})\bigg\}
+\frac{f_G^2m_0^2}{24\pi^2}\int^{t_{f}}_{t_{i}}dt ~G_X^2
\\
&&\qquad +f_G^2m_0^2\left[\frac89\bigg\{g_3^4(t_{f})-g_{3}^4(t_{i})\bigg\}
-\frac{3}{2}\bigg\{g_2^4(t_{f})-g_{2}^4(t_{i})\bigg\}
-\frac{1}{198}\bigg\{g_1^4(t_{f})-g_{1}^4(t_{i})\bigg\}\right] ,
\nonumber  \\
&&m_{u_3^c}^2(t_{f})=m_{u_3^c}^2(t_{i})+\frac13\bigg\{X_t(t_{f})-X_t(t_{i})\bigg\}
+\frac{f_G^2m_0^2}{12\pi^2}\int^{t_{f}}_{t_{i}}dt ~G_X^2
\\
&&\qquad +f_G^2m_0^2\left[\frac89\bigg\{g_3^4(t_{f})-g_{3}^4(t_{i})\bigg\}
-\frac{8}{99}\bigg\{g_1^4(t_{f})-g_{1}^4(t_{i})\bigg\}\right] ,
\nonumber \\
&&m_{h_u}^2(t_{f})=m_{h_u}^2(t_{i})+\frac12\bigg\{X_t(t_{f})-X_t(t_{i})\bigg\}
+\frac{f_G^2m_0^2}{8\pi^2}\int^{t_{f}}_{t_{i}}dt ~G_X^2
\\
&&\qquad -f_G^2m_0^2\left[\frac32\bigg\{g_2^4(t_{f})-g_{2}^4(t_{i})\bigg\}
+\frac{1}{22}\bigg\{g_1^4(t_{f})-g_{1}^4(t_{i})\bigg\}\right] ,
\nonumber
\end{eqnarray}
where 
\dis{
&\quad\quad~ X_t(t_{f})-X_t(t_{i})=X_t(t_{i})\left[e^{\frac{3}{4\pi^2}\int^{t_{f}}_{t_i}dt ~y_t^2}-1\right]
\\
&+e^{\frac{3}{4\pi^2}\int^{t_{f}}_{t_i}dt ~y_t^2}
\int^{t_{f}}_{t_i}dt^\prime\left(\frac{3}{4\pi^2}y_t^2A_t^2-\frac{f_G^2m_0^2}{4\pi^2}G_X^2\right)
e^{\frac{-3}{4\pi^2}\int^{t^\prime}_{t_i}dt^{\prime\prime}y_t^2}
}
and 
\dis{
A_t(t_f)=e^{\frac{3}{4\pi^2}\int^{t_{f}}_{t_i}dt ~y_t^2}
\left[A_t(t_i) 
-\frac{f_Gm_0}{8\pi^2}\int^{t_{f}}_{t_i}dt^\prime 
G_Ae^{\frac{-3}{4\pi^2}\int^{t^\prime}_{t_i}dt^{\prime\prime} ~y_t^2}\right] .
}
In the main text, we set $t_i=t_M$, $t_f=t$, 
and define $F_2(t)$ as  
\dis{ 
&\qquad F_2(t)\equiv e^{\frac{3}{4\pi^2}\int^{t}_{t_M}dt^\prime y_t^2} \int^{t}_{t_M}dt^\prime ~\frac{3}{4\pi^2}y_t^2A_t^2 ~e^{\frac{-3}{4\pi^2}\int^{t'}_{t_M}dt^{\prime\prime}y_t^2}
\\
&-\frac{f_G^2m_0^2}{4\pi^2} \left[e^{\frac{3}{4\pi^2}\int^{t}_{t_M}dt^\prime y_t^2} \int^{t}_{t_M}dt^\prime ~G_X^2 ~e^{\frac{-3}{4\pi^2}\int^{t'}_{t_M}dt^{\prime\prime}y_t^2}
-\int^{t}_{t_M}dt^\prime~G_X^2 \right] .
}


\end{document}